\documentclass[twocolumn]{aastex63}

\usepackage{graphicx}	
\usepackage{color}
\usepackage{amsmath}	% Advanced maths commands
\usepackage{amssymb}	% Extra maths symbols
\usepackage{mathrsfs}
\usepackage{mathrsfs}
\usepackage{multirow}

\usepackage{makecell}
\usepackage{booktabs}
\usepackage{array}

\newcommand{\msun}{M_\odot}
\newcommand{\msunyr}{M_\odot~{\rm yr}^{-1}}
\newcommand{\mpc}{\rm cMpc}
\newcommand{\Muv}{M_{\rm UV}}

\newcommand{\HbOIII}{H$\beta+$[\ion{O}{3}] }

\shorttitle{$z=5$ H$\alpha$ and \HbOIII Dual-line Emitters in JWST CEERS Field}
\shortauthors{Guo et al.}

\begin{document}

\title{A Search for $z=5$ H$\alpha$ and \HbOIII Dual-Line Emitting Galaxies in the JWST CEERS Field: Implications for the AGN Abundance}

\author[0009-0002-6578-8110]{Jingsong Guo}
%\email{2100011602@stu.pku.edu.cn}
\affiliation{Department of Astronomy, School of Physics, Peking University, Beijing 100871, China}

\author[0000-0003-2984-6803]{Masafusa Onoue}
\email{masafusa.onoue@ipmu.jp}
\altaffiliation{Kavli Astrophysics Fellow}
\affiliation{Kavli Institute for the Physics and Mathematics of the
Universe (Kavli IPMU, WPI), The University of Tokyo Institutes for Advanced Study, The University of Tokyo, Kashiwa, Chiba 277-8583, Japan}
\affiliation{Center for Data-Driven Discovery, Kavli IPMU (WPI), UTIAS, The University of Tokyo, Kashiwa, Chiba 277-8583, Japan}
\affiliation{Kavli Institute for Astronomy and Astrophysics, Peking
University, Beijing 100871, China}

\author[0000-0001-9840-4959]{Kohei Inayoshi}
\email{inayoshi@pku.edu.cn}
\affiliation{Kavli Institute for Astronomy and Astrophysics, Peking University, Beijing 100871, China}

\author[0000-0002-8360-3880]{Dale D. Kocevski}
\affiliation{Department of Physics and Astronomy, Colby College, Waterville, ME 04901, USA}

\author[0000-0001-8519-1130]{Steven L. Finkelstein}
\affiliation{Department of Astronomy, The University of Texas at Austin, Austin, TX, USA}

\author[0000-0002-9921-9218]{Micaela B. Bagley}
\affiliation{Department of Astronomy, The University of Texas at Austin, Austin, TX, USA}

\author[0000-0001-8688-2443]{Elizabeth J. McGrath}
\affiliation{Department of Physics and Astronomy, Colby College, Waterville, ME 04901, USA}

\begin{abstract}
The James Webb Space Telescope (JWST) has enabled us to uncover faint galaxies and active galactic nuclei (AGNs) in the early universe. Taking advantage of the unique filter combination used in the Cosmic Evolution Early Release Science Survey (CEERS) program, we perform an extensive photometric search of galaxies emitting strong \HbOIII and H$\alpha$ lines.
The redshift range of the galaxies is limited to $5.03\leq z\leq 5.26$ by requiring photometric excesses in NIRCam's F277W and F410M images.
A total of 261 \HbOIII and H$\alpha$ dual-line emitters are found over the absolute UV magnitude  $-22\lesssim M_{\mathrm{UV}}\lesssim -17$, with a mean rest-frame equivalent width of 1010 \AA\ for \HbOIII\ and 1040 \AA\ for H$\alpha$.
This population accounts for $\sim 40\%$ of the Lyman break galaxies at this redshift range. 
Intriguingly, there are 58 objects (22\% of the whole sample) that exhibit compact morphology at the rest-UV or optical wavelength. 
With an assumption that these compact dual-line emitters are dominated by AGN, their AGN bolometric luminosities are in the range of $2\times 10^{43} \lesssim L_{\rm bol}/({\rm erg~s}^{-1})\lesssim 3\times 10^{44}$.
Their number density is two orders of magnitude higher than the extrapolation from the UV-selected luminous quasars, which is in good agreement with previous JWST studies of broad-line AGNs, requiring a $\sim 10\%$ of the AGN duty cycle. 
%These objects have \HbOIII emission in F277W and H$\alpha$ in F410M, and their redshifts thus are restricted to $5.03\leq z\leq 5.26$. 
%Two of these dual-line emitters have been spectroscopically confirmed by existing JWST data.
%These dual-line emitters have an absolute UV magnitude range of $-22\lesssim M_{\mathrm{UV}}\lesssim -17$ and a mean rest-frame equivalent width of 1010 \AA\ for \HbOIII\ and 1040 \AA\ for H$\alpha$. We find that this population accounts for $\sim 40\%$ of the Lyman break galaxies at this redshift range. 
%Assuming that the compact dual-line emitters are dominated by AGNs, we find that the abundance of these unobscured AGN candidates with the bolometric luminosity range of $2\times 10^{43} \lesssim L_{\rm bol}/({\rm erg~s}^{-1})\lesssim 3\times 10^{44}$ ($\simeq 10^{-3}~{\rm cMpc}^{-3}$) is two orders of magnitude higher than the extrapolation from the UV-selected luminous quasars.
%This high abundance is yet comparable to those of broad-line AGNs, requiring a $\sim 10\%$ of the AGN duty cycle of their host galaxies. 
%This result is in good agreement with previous JWST studies of broad-line AGNs, requiring a $\sim 10\%$ of the AGN duty cycle. 
%The black hole mass function constructed by assuming the median Eddington ratio for JWST-identified AGN samples covers $M_{\rm BH}\simeq 10^6 - 10^7~\msun~(\langle \lambda_{\rm Edd}\rangle/0.3)^{-1}$, bridging the gap between their seed population and those previously observed with ground-based telescopes.
Moreover, our dual-line emitter sample reaches the faint end of the H$\alpha$ and [\ion{O}{3}] luminosity functions down to $\lesssim 10^{42}~{\rm erg~s}^{-1}$. 
Spectroscopic follow-up observations are planned in an approved JWST Cycle~3 program, in which we aim to confirm their nature, characterize their black hole activity, and construct their mass distribution at $10^6\lesssim M_{\rm BH}/\msun \lesssim 10^8$.

\end{abstract}

\section{Introduction} \label{sec:intro}

One of the most remarkable discoveries from the first year of observations by the James Webb Space Telescope (JWST; \citealt{Rigby_2022}) is the substantial detection of low-luminosity active galactic nuclei (AGN) in the early universe.
While hundreds of quasars were reported down to $z=7.6$  in the pre-JWST era \citep{Wang_2021, Fan_2023},
these quasars are limited to bolometric luminosity $L_{\rm bol}\gtrsim 10^{46}\ \mathrm{erg\ s^{-1}}$, or supermassive black hole (SMBH) mass of  $M_\mathrm{BH}\gtrsim 10^8 M_\odot$.
Therefore, rapidly accreting low-mass SMBHs, which are a key population for understanding the early phase of SMBH growth, have been largely missing.

The unprecedented sensitivity of the JWST in infrared wavelengths has transformed the situation.
The Cycle~1 observations of the JWST have identified a number of broad-line AGN at $z>4$ \citep[e.g.,][]{Onoue_2023,Kocevski_2023, Kocevski_2024, Harikane_2023_agn,Maiolino_2023_JADES,Matthee_2024,Larson_2023,Greene_2024,Treiber_2024_UNCOVERAGN}, followed by photometric candidates  \citep{Kocevski_2024,  Kokorev_2024, Akins_2024LRD}.
Some of the earliest galaxies are also claimed to host AGN down to $z\sim11$  \citep{Maiolino_2023, Goulding_2023, Bogdan_2024}.
These high-redshift JWST AGN generally cover a luminosity range $\approx2$--3 dex lower than that of the ground-based sample.
Broad Balmer line detection of these JWST AGNs also enables virial BH mass estimates, yielding to discoveries of $M_\mathrm{BH}\sim 10^6 M_\odot$ SMBHs \citep{Maiolino_2023_JADES}.

The unexpectedly high abundance of these JWST AGNs is well manifested in the AGN luminosity function (LF), a fundamental quantity that can be constrained from AGN surveys \citep[e.g.,][]{Niida_2020, Matsuoka_2018, Matsuoka_2023, Schindler_2023}.
The early JWST discoveries of broad-line AGNs show that the faint end of the  $z>5$ UV LF is significantly ($\gg1$ dex) higher than what is expected from the extrapolation of the ground-based studies that cover a higher luminosity range \citep{Kocevski_2023, Harikane_2023_agn, Matthee_2024, Maiolino_2023_JADES, Greene_2024}.
The excess is also present but somewhat milder in the H$\alpha$ LF \citep{Matthee_2024, Greene_2024}. 
While their nature, especially those with red continua in the rest optical is still controversial \citep{Kokubo_2024, Li_LRD_2024, WangB_2024, Baggen_2024}, the apparent high AGN abundance at the faint end may affect the discussion on the AGN ionizing photon budget driving 
cosmic reionization \citep{Matsuoka_2018, Giallongo2019, Yung_2021} and the rapid mass assembly of SMBHs at high redshift \citep[e.g.,][]{Trinca_2022, %Li_2023,
Li_LF_2024, Schneider_2023}. 
However, it should be pointed out that these studies are based on samples of a few dozen AGN, and their sample selections are heterogeneous across different studies.
Therefore, a larger sample of high-redshift low-luminosity AGN with a clear selection function is awaited.

Here we perform a search for dual-line emitters of \HbOIII\ and H$\alpha$ at $z\sim5$, making use of the NIRCam photometric data of the Cosmic Evolution Early Release Science Survey (CEERS; PI: S. Finkelstein). 
The CEERS survey provides 7-band JWST/NIRCam photometry in 10 pointings (97 arcmin$^2$) with 5$\sigma$ depths of 28.4 -- 29.2 magnitude \citep{Bagley_CEERS_2023}. 
The relatively narrow bandpasses of mediumband filters enable us to capture strong line emission \citep{davis2023census, Withers23, Williams_2023_JEMS}, as is also predicted in data simulations \citep{wilkins2023cosmic}.
Moreover, JWST spectroscopic studies have unveiled a wealth of \HbOIII\ emitters at high redshift \citep{Kashino_2023, Matthee_2023_EIGERII, Meyer_2024}.
Such strong \HbOIII\ emission can be detected in broadband SED analyses.

There are several advantages to using dual-line emitter selection for high-redshift sources.
First, it allows for the selection of line emitters within a narrow redshift range, especially when using mediumband or narrowband filters.
Additionally, it also helps suppress contamination from foreground and background emission line galaxies by requiring photometric excess in two different filters (e.g., F277W for \HbOIII\ and  F410M for H$\alpha$ at $z\sim5.2$).
Moreover, the emitter selection is effective down to the limiting magnitudes of the continuum filters, while the commonly-used Lyman-break technique is usually limited by the depths of the filters blueward of the redshfited Lyman break in order to observe sufficient color excesses between the broadband filters around the Lyman break.

\begin{figure*}
\centering
\includegraphics[width=140mm]{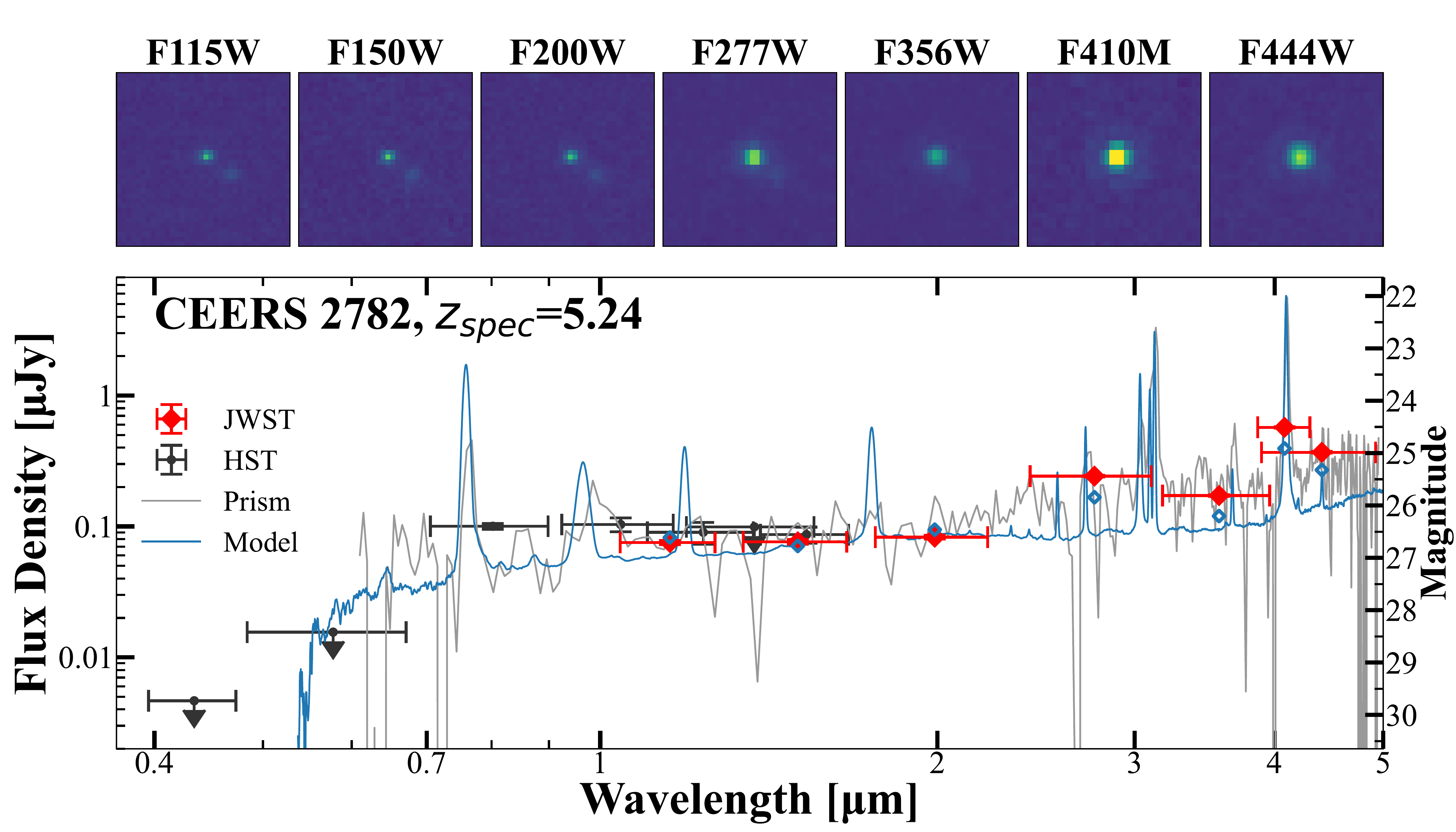}
\caption{A spectroscopically-confirmed broad-line AGN at $z_\mathrm{spec}=5.24$ \citep[CEERS 2782,][]{Onoue_2023, Kocevski_2023}.
(\textit{Top:}) The $1''.5 \times 1''.5$ postage stamps of the NIRCam images.
The filter name for each image is indicated at the top of each panel.
(\textit{Bottom:}) The rest-frame UV-to-optical SED. The JWST/NIRCam and HST photometry is shown in red and black, respectively.
The wavelength coverage of each filter is indicated by the errorbar.
The upper limits show $1\sigma$ detection limits.
The NIRSpec PRISM spectrum presented in \citet{Kocevski_2023} is shown in grey, which is scaled to match the NIRCam photometry.
The redshifted and modified version of the composite spectrum of low-redshift quasars \citep{VandenBerk_2001} is shown in cyan (see Section~\ref{sec:Model} for details).
The expected NIRCam photometry convolved from the model is shown as open cyan diamonds, demonstrating that our model can well reproduce the real photometry measurement.
}
\vspace{5mm}
\label{fig:confirmedAGN}
\end{figure*}

We initiate this work motivated by early high-redshift AGN studies of \citet{Onoue_2023} and \citet{Kocevski_2023}, where they discovered a $z=5.24$ AGN, CEERS 2782, from a catalog of Lyman-break galaxies \citep{Bouwens_2021}.
Figure \ref{fig:confirmedAGN} shows the observed images and SEDs of CEERS 2782, in which one can see a strong emission line excess of \HbOIII\ in F277W and H$\alpha$ in F410M (and partly in F444W as well).
CEERS 2782 was selected as a compact line emitter with the rest-frame equivalent width $\mathrm{EW_{H\beta+[O III]}}=1100\ \mathrm{\AA}$ and $\mathrm{EW_{H\alpha}}=1600$\ \AA, later confirmed as a broad-line AGN by a follow-up JWST/NIRSpec MSA spectroscopy.
This work is an extension of these early works, aiming at constructing a large line emitter catalog at $z\sim5$ in the CEERS field and selecting AGN candidates to constrain the number statistics of low-luminosity AGN at $z\sim5$.

This paper is organized as follows. In Section \ref{sec:method}, we introduce our sample and method for selecting dual-line emitters and AGN candidates. Section \ref{sec:results} describes the properties of dual-line emitters, while in Section \ref{sec:LF} we discuss the luminosity functions of them. We finally summarize this work in Section \ref{sec:conclusion}.

%%%%%%%%
%	Fig.1.     %
%%%%%%%%

Throughout the paper, we assume a flat $\Lambda$CDM Universe with $\mathrm{H_0}=70~\mathrm{km\ s^{-1}\ Mpc^{-1}}$ and $\Omega_\mathrm{m}=0.3$.
Magnitudes quoted in this paper are in the AB system \citep{Oke_Gunn_1983}.

\section{Method} \label{sec:method}
\subsection{Sample} \label{sec:samples}

The parent sample of this study is a publicly available JWST photometric catalog in the CEERS field with 7-band NIRCam data (F115W, F150W, F200W, F277W, F356W, F410M and F444W), provided by the DAWN JWST Archive (DJA).
The NIRCam imaging data was processed by the \textsc{grizli} pipeline \citep{brammer_2023_grizli}. 
The pixel scale of the stacked images is $0\arcsec.04$ for all filters.
The catalog also contains 7-band HST photometry result (F435W, F606W, F814W, F105W, F125W, F140W and F160W) which data is provided by CANDELS and UVCANDELS projects \citep{Grogin_2011_CANDELS,Koekemoer_2011_candels}.
We also use the NIRSpec spectra in the same fields from the DJA archive, which are reduced by \textsc{msaexp} \citep{brammer_2023_msaexp}.
More details of the NIRCam and NIRSpec data products are described in \citet{Valentino_2023} and \citet{heintz2023extreme}.

%%%%%%%%
%	Fig.2.     %
%%%%%%%%
\begin{figure*}
\begin{center}
{\includegraphics[width=85mm]{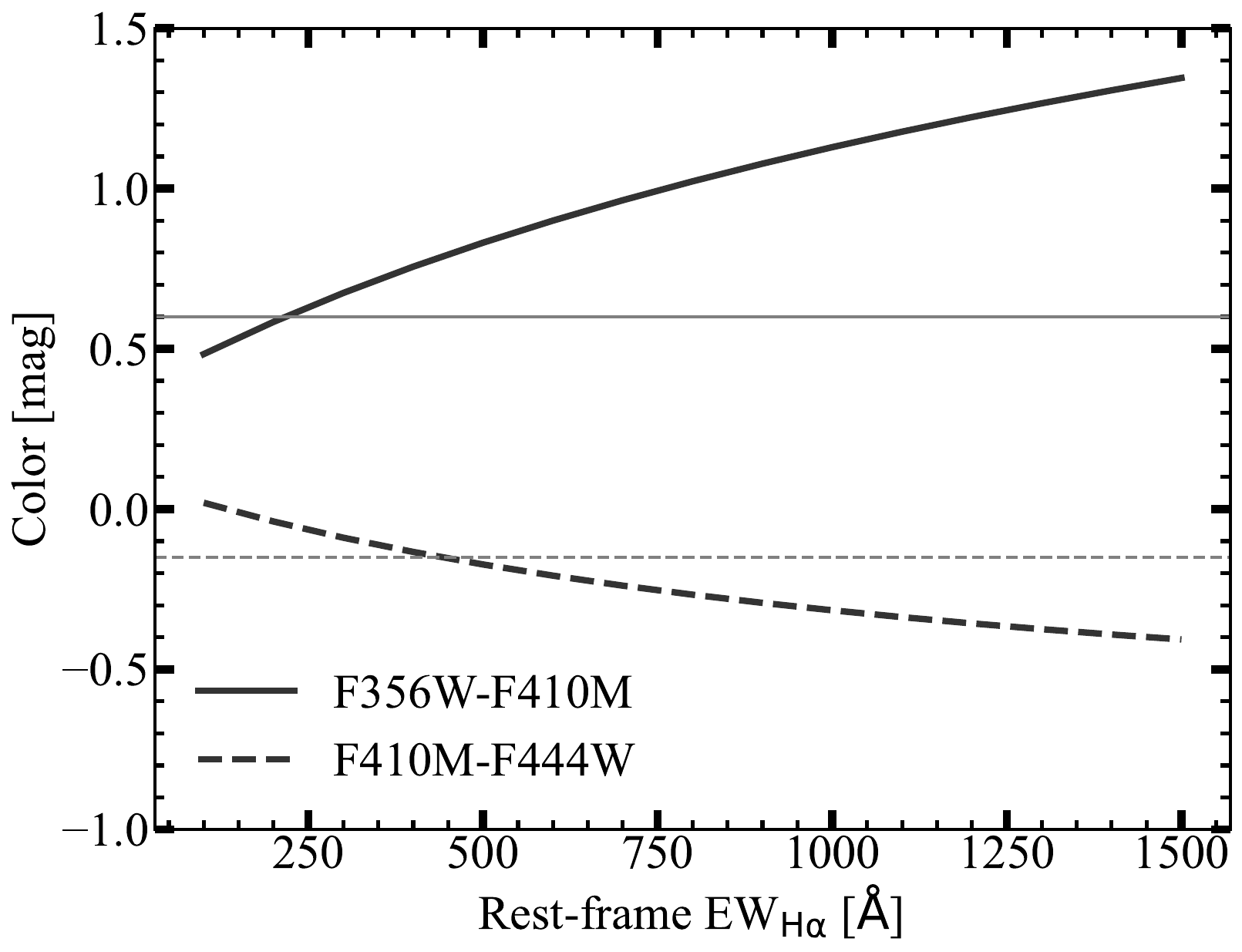}}\hspace{5mm}
{\includegraphics[width=85mm]{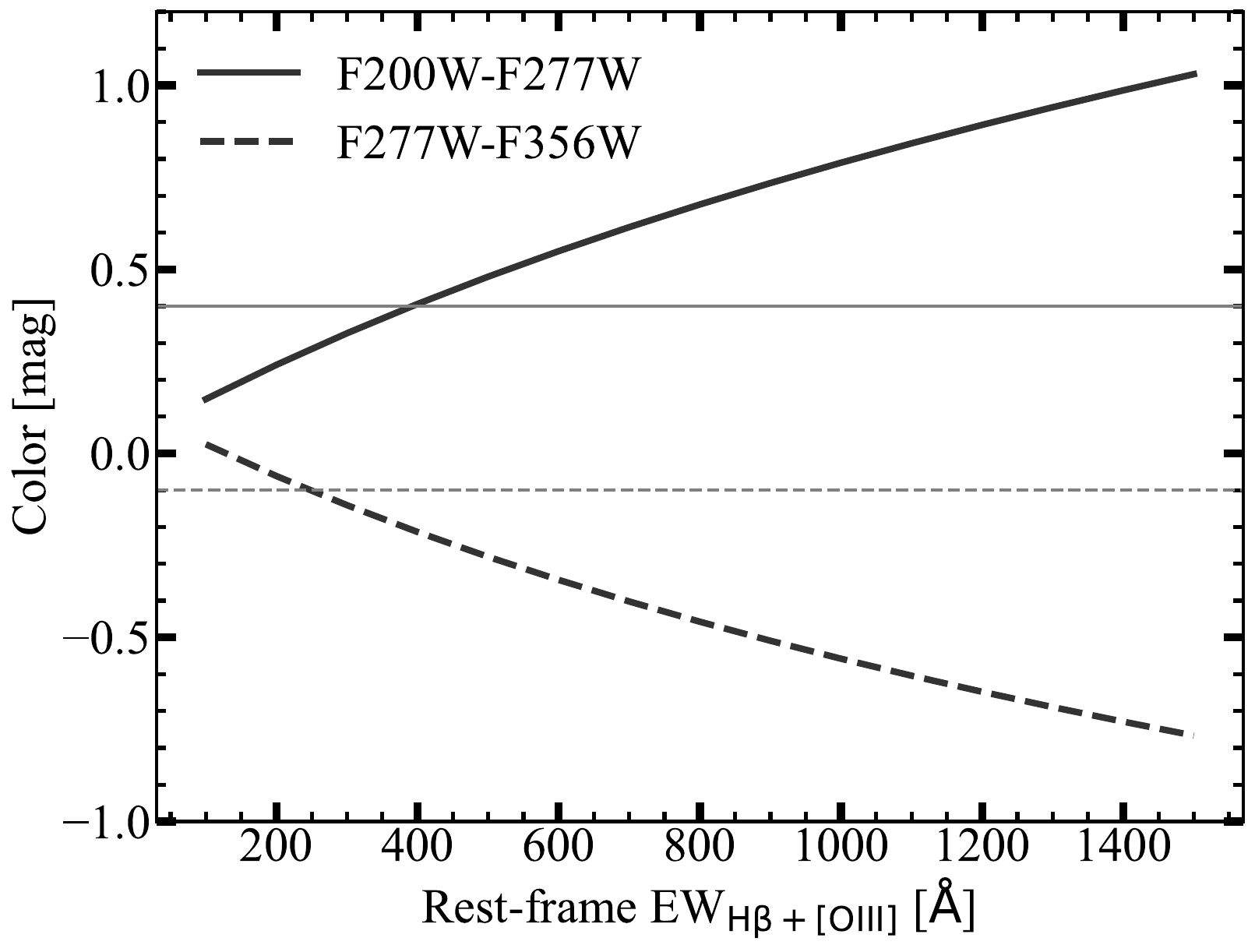}}
\caption{The JWST/NIRCam colors of a line emitter at $z=5.2$ as a function of rest-frame equivalent widths of H$\alpha$ (left) and \HbOIII (right).
We present colors of F356W $-$ F410M and F410M $-$ F444W colors for H$\alpha$, while F200W $-$ F277W and F277W $-$ F356W for H$\beta+$[\ion{O}{3}].
We assume a single power-law continuum with $\alpha_\lambda =-1.5$, following that of a typical quasar \citep{VandenBerk_2001}.
The color-cut thresholds we use to select $z=5.2$ dual-line emitters are indicated by horizontal lines (see Section~\ref{sec:selection}).
}\label{fig:excess}
\end{center}
\end{figure*}

\subsection{Photometric Excess} \label{sec:excess}

The advantage of utilizing strong emission lines of \HbOIII and H$\alpha$ in photometric selection is that one can suppress contamination of low- or high-redshift interlopers without relying on the Lyman break selection technique. 
Moreover, this dual-line emitter selection enables the identification of line emitters down to the continuum sensitivity limit of the JWST.
To illustrate the effects of these strong emission lines on the photometric colors of NIRCam broad/medium band filters, 
Figure~\ref{fig:excess} shows the NIRCam colors of a $z=5.2$ line emitter as a function of H$\alpha$ and \HbOIII equivalent widths (EWs)
in the rest frame. 
We note that at this particular redshift, the H$\alpha$ and \HbOIII multiplet enter the F410M and F277W filter, respectively.
Here, we consider a single emission line (or a multiplet) on top of a single power-law continuum with an index of 
$\alpha_\lambda (\equiv {\rm d}\ln F_\lambda /{\rm d}\ln \lambda)=-1.5$, consistent with those for low-redshift quasars \citep{VandenBerk_2001}.
We assume the EW ratio of the \HbOIII\ multiplet as $ \mathrm{EW_{H\beta}}: \mathrm{EW_{[OIII]\lambda4960 }} : \mathrm{EW_{[OIII]\lambda5007 }} = 0.3: 0.25: 1$, based on the measurement for line-emitting galaxies at $z\sim 1.7$ -- $6.7$ \citep{Withers23}.
The line widths of the H$\alpha$ and H$\beta$ emission are assumed to be $4,000$ km s$^{-1}$ and the [\ion{O}{3}] doublet to $300$ km s$^{-1}$, respectively.

The left panel of Figure~\ref{fig:excess} shows the response of F410M as the EW$_\mathrm{H\alpha}$ varies from 100 \AA\ to 1500 \AA.
The F356W$-$F410M color significantly increases from 0.5 to 1.4 due to the relatively narrow bandpass of F410M.
Such a large photometric excess can be captured by color selection for galaxies detected by JWST.
The response of the F410M$-$F444W color is also significant (ranging from $0.0$ to $-0.4$) but milder because F444W overlaps with the wavelength coverage of F410M.

The right panel of Figure~\ref{fig:excess} shows the response of the broadband F277W filter, which presents 
a similar behavior with F200W$-$F277W increasing up to $1.0$ mag and F277W$-$F356W decreasing down to $-0.8$ mag.
As we discuss in Section~\ref{sec:selection}, we employ color cuts to select dual-line emitters at $z\simeq 5.2$ 
based on these colors.

\subsection{SED Models}\label{sec:Model}

To determine the color selection for $z=5.2$ H$\alpha$ and \HbOIII dual-line emitters,
we model the SEDs of line-emitting AGNs and galaxies.
We adopt an AGN SED template based on the composite spectrum for low-redshift luminous quasars presented in \citet{VandenBerk_2001}.
This SED template covers rest-frame wavelengths of $\lambda_\mathrm{rest}=800-8555$ \AA. 
The rest-frame EWs in this SED template are $\mathrm{EW_{H\alpha}}= 194~{\rm \AA}$ and $\mathrm{EW_{H\beta+[OIII]}}= 63 ~{\rm \AA}$.
This model is connected to another AGN SED template \citep{Glikman_2006} that covers rest-frame near-infrared wavelengths up to $3.5 ~\micron$. We note that the luminosity range of these two AGN templates are comparable, so that our connection is appropriately applied. Therefore, the wavelength coverage of the SED model, from rest-frame ultraviolet to near-infrared bands, enables us to simulate the colors of NIRCam filters from $z\sim 1$ to 10.

%%%%%%%%
%	Fig.3.     %
%%%%%%%%
\begin{figure*}
\begin{center}
{\includegraphics[width=85mm]{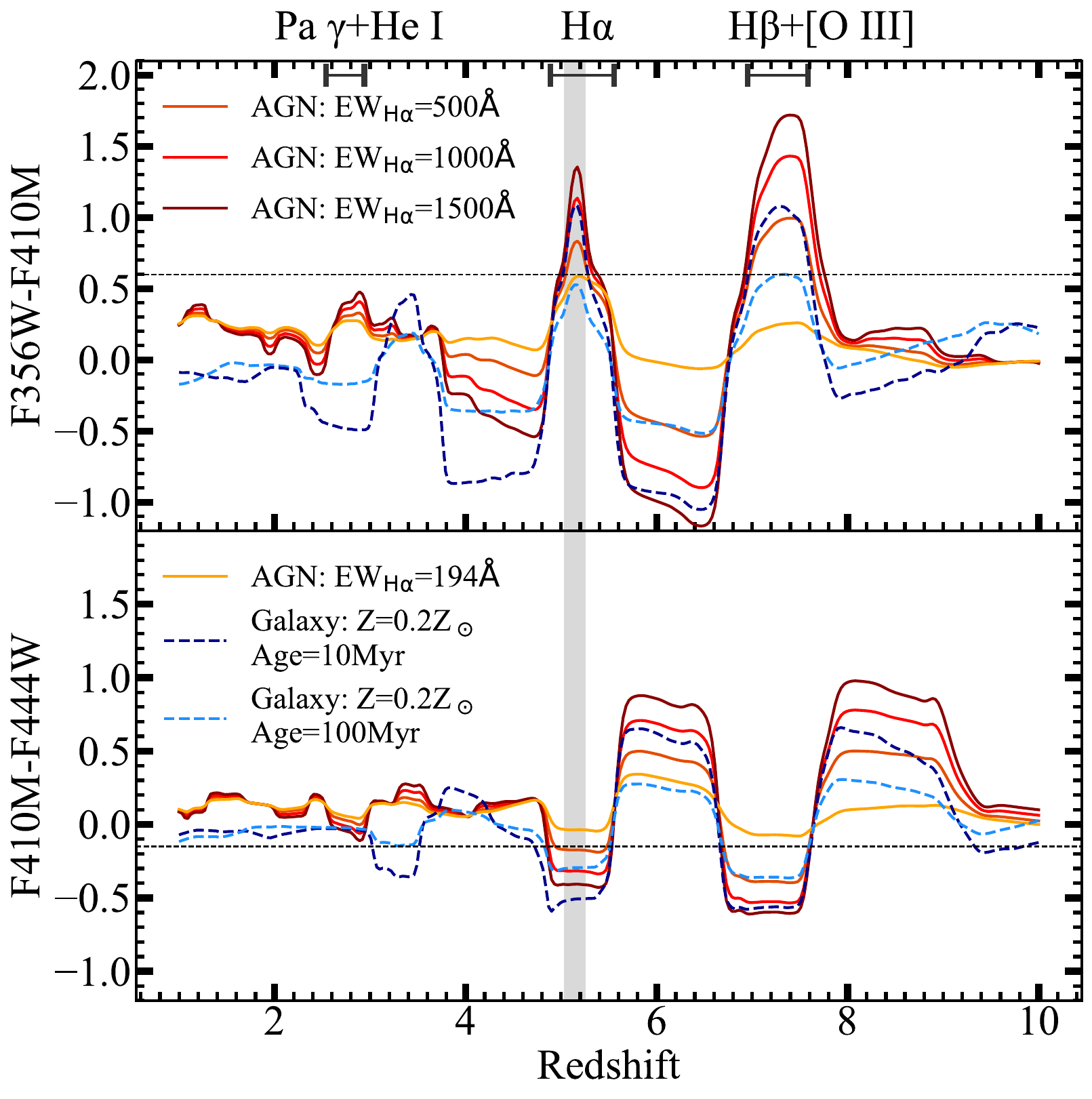}}\hspace{5mm}
{\includegraphics[width=85mm]{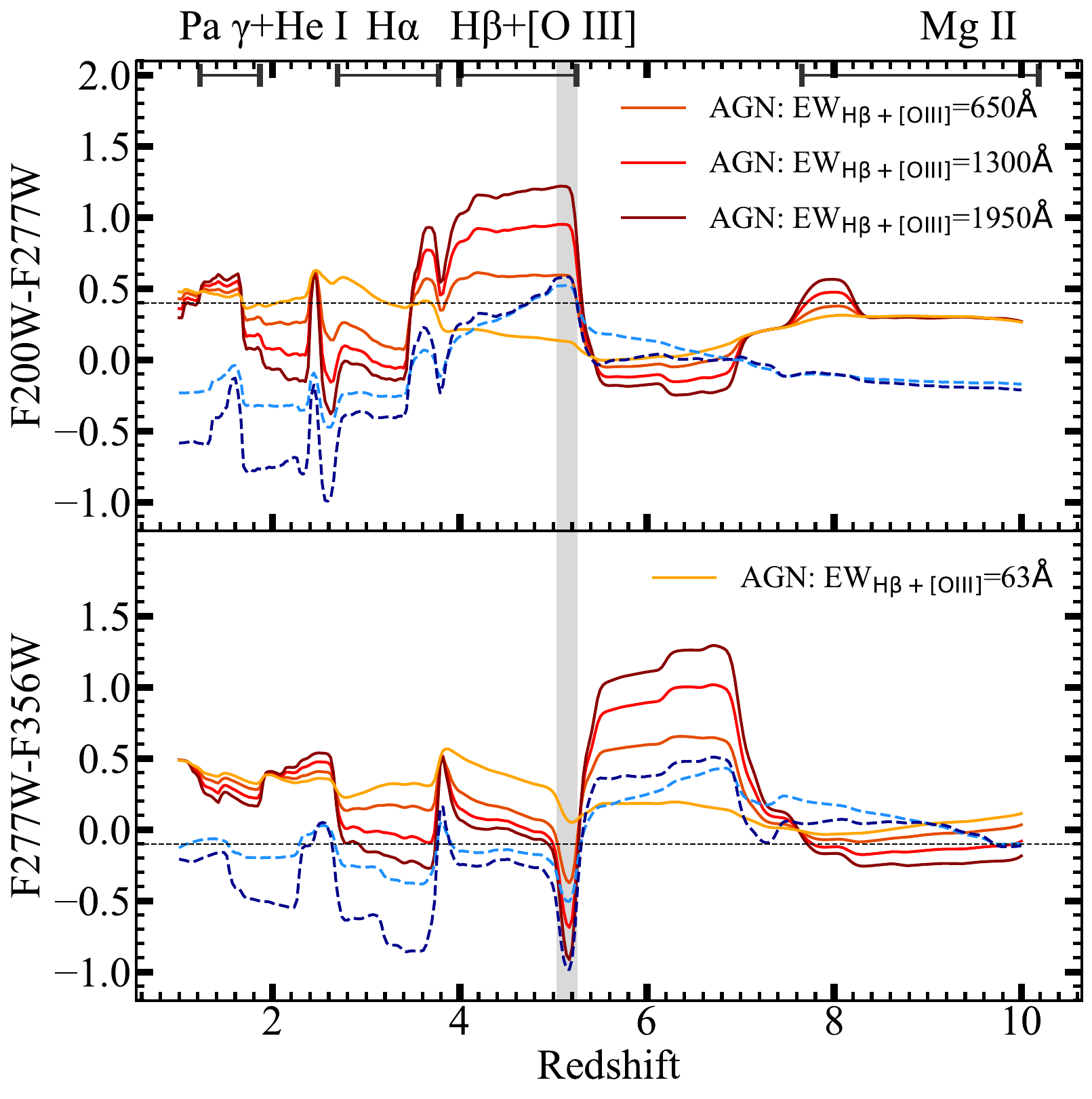}}
\caption{JWST/NIRCam colors of AGNs and galaxies as the function of redshift.
\textit{(Left:)} F356W $-$ F410M (top) and F410M $-$ F444W (bottom).
\textit{(Right:)}  F200W $-$ F277W (top) and F277W $-$ F356W (bottom).
For AGN SED models, we show those with $\mathrm{EW_{H\alpha}}=500~{\rm \AA}$ (dark orange), $1000~{\rm \AA}$ (red), and $1500~{\rm \AA}$ (dark red) (see Section~\ref{sec:selection} for more details).
The same curve for the original quasar SED model is also shown in orange (\citealt{VandenBerk_2001}, \citealt{Glikman_2006}).
For galaxy SED models, we show two different ages of 10 Myr and 100 Myr and with a metallicity of $Z=0.2~Z_\odot$ from \citet{Inoue_2011}.
The redshift window of interest in this work is highlighted by grey shades in both panels. 
On top of the figures, we mark the redshift ranges where important emission lines enter the F410M (\textit{left}) and F277W (\textit{right}).
The horizontal dashed lines show the color thresholds for our dual-line emitter selection (Section~\ref{sec:selection}).
}
\label{fig:fig_1_Ha}
\end{center}
\end{figure*}

It is empirically known that faint AGN have strong emission lines with respect to their continuum \citep{Baldwin77}.
In order to implement the so-called Baldwin effect, we refer to \citet{Withers23}, in which strong H$\alpha$ and \HbOIII 
dual-line emitters are selected from JWST mediumband images over $1.7\lesssim z \lesssim 6.7$.
The median EWs of these line emitters are $\mathrm{EW_{H\alpha}}= 893$ \AA\ and $\mathrm{EW_{H\beta+[OIII]}}= 1255$ \AA.
Motivated by this result, we construct our AGN SED model by assuming the EW ratio of
\begin{equation}\label{eq:EWratio}
\mathrm{EW_{H\alpha}}: \mathrm{EW_{H\beta+[OIII]}} = 1: 1.3,     
\end{equation}
and scaling the EW$_{\rm H\alpha}$ from 200 \AA\ to 1500 \AA.
The upper limit of the EW$_{\rm H\alpha}$ is chosen to match the observed JWST AGN (\citealt{Onoue_2023}).
As discussed later, our color selection criteria are designed to effectively identify line emitters with H$\alpha$ EWs of $\gtrsim 500$~\AA. 
Therefore, the choice of the maximum EW value does not impact the threshold of our color cuts (see Section~\ref{sec:selection}). 
Other emission lines at $\lambda_\mathrm{rest}>1215\ {\rm \AA}$ that have rest-frame $\mathrm{EW}>3$ \AA\ listed 
in \citet{VandenBerk_2001} and \citet{Glikman_2006} are scaled with the EW of H$\alpha$ line, except for the pseudo-continuum of 
Fe~{\sc ii} and Fe~{\sc iii}.

We also take into account SED models of young star-forming galaxies that exhibit strong Balmer and [\ion{O}{3}] emission lines,
in comparison to the AGN SED with prominent emission lines.
For this purpose, we use the galaxy SED models provided by \citet{Inoue_2011}. 
Their models consider a constant star formation rate history assuming a Salpeter initial mass function \citep{Salpeter_1955} 
with a mass range of $1-100~\msun$ and grids of ages and metallicity. 
For all the models, the gas and stellar metallicity are set to be equal.
For our analysis, we adopt models with $Z=0.2~Z_\odot$ and ages of 10 and 100 Myr.
The metallicity value is chosen so that the [\ion{O}{3}] EW is maximized due to the hotter nebular gas temperature resulting from 
the lack of metal coolants. 
Additionally, we assume that all the ionizing stellar radiation is absorbed in the nebular region and reprocessed into emission lines, allowing us to observe the maximum contribution of these emission lines in our color selection.
Under these metallicity and age conditions, the SED contains strong Balmer and [\ion{O}{3}] emission line simultaneously, 
with a continuum with a UV slope of $-2.8 \lesssim \alpha_\lambda \lesssim -2.3$ (see Figure~11 in \citealt{Inoue_2011}), 
which is bluer than the AGN templates mentioned above.

\subsubsection{F410M for H$\alpha$} \label{sec:excess_Ha}

The left panel of Figure~\ref{fig:fig_1_Ha} shows the redshift dependence of the F356W$-$F410M and F410M$-$F444W colors from $z=1$ to 10.
Here, we present three AGN models with different rest-frame H$\alpha$ EWs of ${\rm EW}_{\rm H\alpha} =500$~\AA, $1000$~\AA, and $1500$~\AA.
The \HbOIII\ EWs of these models are ${\rm EW}_{\rm H\beta+OIII} = 650$~\AA, 1300~\AA, and 1950~\AA, respectively (see Equation~\ref{eq:EWratio}).
The color evolution of the original quasar template is also shown (orange; \citealt{VandenBerk_2001} and \citealt{Glikman_2006}).
Additionally, we include the young galaxy models with $Z=0.2~Z_\odot$ and ages of 10 and 100 Myr from \citet{Inoue_2011}.

As seen in Figure~\ref{fig:excess}, the responses of the F356W$-$F410M and F410M$-$F444W colors are significant when H$\alpha$ emission falls within the F410M wavelength coverage at $z=4.9 - 5.5$.
Within this redshift range, the F410M$-$F444W color becomes blue ($<0$). The F356W$-$F410M color becomes red when neither H$\alpha$ nor \HbOIII enter F356W, effectively limiting the redshift window to $z=5.03 - 5.26$.

On the other hand, strong \HbOIII\ emission can also achieve a similar F410M excess at $z=6.9 - 7.6$.
Other minor color excesses are observed at $z\sim1.2$ and $z\sim2.8$, when \ion{He}{1} $\lambda$20580 and 
\ion{He}{1} $\lambda$10830+Pa$\gamma$ enter the F410M filter, respectively. 
Therefore, the color selection based on a single emission line can introduce significant contamination from low- or high-redshift interlopers.

\subsubsection{F277W for \HbOIII} \label{sec:excess_O3}
The right panel of Figure~\ref{fig:fig_1_Ha} shows that the F200W$-$F277W and F277W$-$F356W broadband colors 
respond to strong emission lines, while the color excess is milder than those seen in the colors associated with 
the F410M medium band filter.
The color evolution is more complicated than that for H$\alpha$ emitters in F410M.
The F200W$-$F277W color becomes red when H$\alpha$ falls within F277W and \HbOIII falls outside of F200W, occurring at $z=3.44 - 3.77$. 
The F200W$-$F277W color becomes red again when \HbOIII enters F277W at $z=3.83-5.26$, during which F200W is 
free from strong emission lines and thus traces continuum emission. 
Additionally, \ion{Mg}{2} $\lambda 2798$ contributes to a red F200W$-$F277W color at $z\sim 8$.
On the other hand, the F277W$-$F356W color becomes blue at $z=5.03$--$5.26$, when neither \HbOIII nor H$\alpha$ enters F356W.

\subsection{Color criteria for dual-line emitter selection} \label{sec:selection}

We select H$\alpha$ and \HbOIII dual-line emitters at  $z=5.03$--$5.26$ with the following color criteria:
\begin{align}
{\rm F356W}-{\rm F410M}&>0.60 \nonumber\\
\&~{\rm F410M}-{\rm F444W}&<-0.15,\label{eq:Ha}
\end{align}
for the H$\alpha$ line, and 
\begin{align}
{\rm F200W}-{\rm F277W}&>0.40 \nonumber\\
\&~{\rm F277W}-{\rm F356W}&<-0.10 \label{eq:HbO3}
\end{align}
for the \HbOIII multiplet.
These color thresholds are based on our AGN SED model with EW$_{\rm H\alpha}=500$ \AA\ and EW$_{\rm H\beta+[O III]}=650$ \AA.

We also adopt the aperture magnitude with a diameter of $d=0.5\ \mathrm{arcsec}$,
and impose SNR cuts for F277W, F356W, and F410M as
\begin{align}
{\rm SNR}_{\rm F277W}&>5.0, \nonumber\\
{\rm SNR}_{\rm F356W}&>3.0, \nonumber\\
{\rm SNR}_{\rm F410M}&>5.0
\end{align}
in order to select objects that are significantly detected in the H$\alpha$ and \HbOIII\ images.
For the two bands of F200W and F444W that we use in our color selection criteria but do not impose their SNR thresholds, 
we request the 1$\sigma$ upper limit of detection of that band to be consistent with the color selection above, if the SNR of one band is less than 3.

Following these criteria, we expect to detect dual-line emitters with $\mathrm{EW_{H\alpha}}>500~\mathrm{\AA}$ and 
$\mathrm{EW_{H\beta+[O III]}}>650~\mathrm{\AA}$ at $5.03 \leq z \leq 5.26$.
We notice that, at the center of the sensitive redshift range of our method, a dual-line emitter can produce a more significant color excess in F277W compared to the edge of the redshift range (the right pannel of Figure \ref{fig:fig_1_Ha}). 
Thus, though the criteria is set based on $\mathrm{EW_{H\beta+[O III]}}>650~\mathrm{\AA}$, we may also select emitters with weaker \HbOIII at the center of the redshift range.

With the galaxy models, one can expect that star-forming galaxies with moderate metallicity ($0.2~Z_\odot$) and young age ($<100$ Myr) 
are selected if they fall within the redshift range of interest. 
Although the selection criteria are based on the AGN SED model, these conditions essentially capture the emission line features in the photometry.
Therefore, galaxies with strong line emission may also be selected.
It is important to note that the SED models by \citet{Inoue_2011} represent star-forming galaxies that appear as emission line galaxies.
Consequently, the selection of such galaxies is expected if the equivalent widths of their emission lines are sufficiently large, 
which would be the case for young, star-forming galaxies.

%%%%%%%%
%	Table.1.     %
%%%%%%%%
\begin{table}
\renewcommand\thetable{1} %! fix indexing
\caption{The step-by-step summary of our selection}
\label{table:scenario}
\begin{center}
% \tablenum{1}
\begin{tabular}{ccc}
\hline
\hline
Step & Selection & Candidates \\
\hline
1 & All CEERS Objects & 70,514\\[2pt]
2 & SNR$_{\rm F277W}>5$ & 30,675\\
  & SNR$_{\rm F356W}>3$ & \\
  & SNR$_{\rm F410M}>5$ & \\[2pt]
\hline 
\hline
& {\it Dual-Line Emitter Selection}& \\[2pt]
3.1 & F356W$-$F410M $>0.6$   & 817\\
    & F410M$-$F444W $<-0.15$  & \\[2pt]
3.2 & F200W$-$F277W $>0.4$   & 1,150\\
    & F277W$-$F356W $<-0.1$ & \\[2pt]
4 & 3.1+3.2 & 279\\[2pt]
5 & Continuum detection with SNR$>$3 & 275\\
  & in at least two band filters &\\[2pt]
6 & Lyman break and Visual check & {\bf 261}\\
\hline
\hline
& {\it Compact Morphology Selection}& \\[2pt]
7.1 & $R_{\rm e, F115W}<0\arcsec.020$ & 26 \\
7.2 & $R_{\rm e, F356W}<0\arcsec.058$ & 60 \\[2pt]
8.1 & 7.1+7.2 & 18 \\[2pt]
8.2 & 7.1 or 7.2  & 68 \\
9 & 8.2+Visual Check & {\bf 58}\\ \hline
\end{tabular}
\label{tab:jwst_mag}
\end{center}
\end{table}

%%%%%%%%
%	Fig.4.     %
%%%%%%%%
\begin{figure*}  
\begin{center}
{\includegraphics[width=85mm]{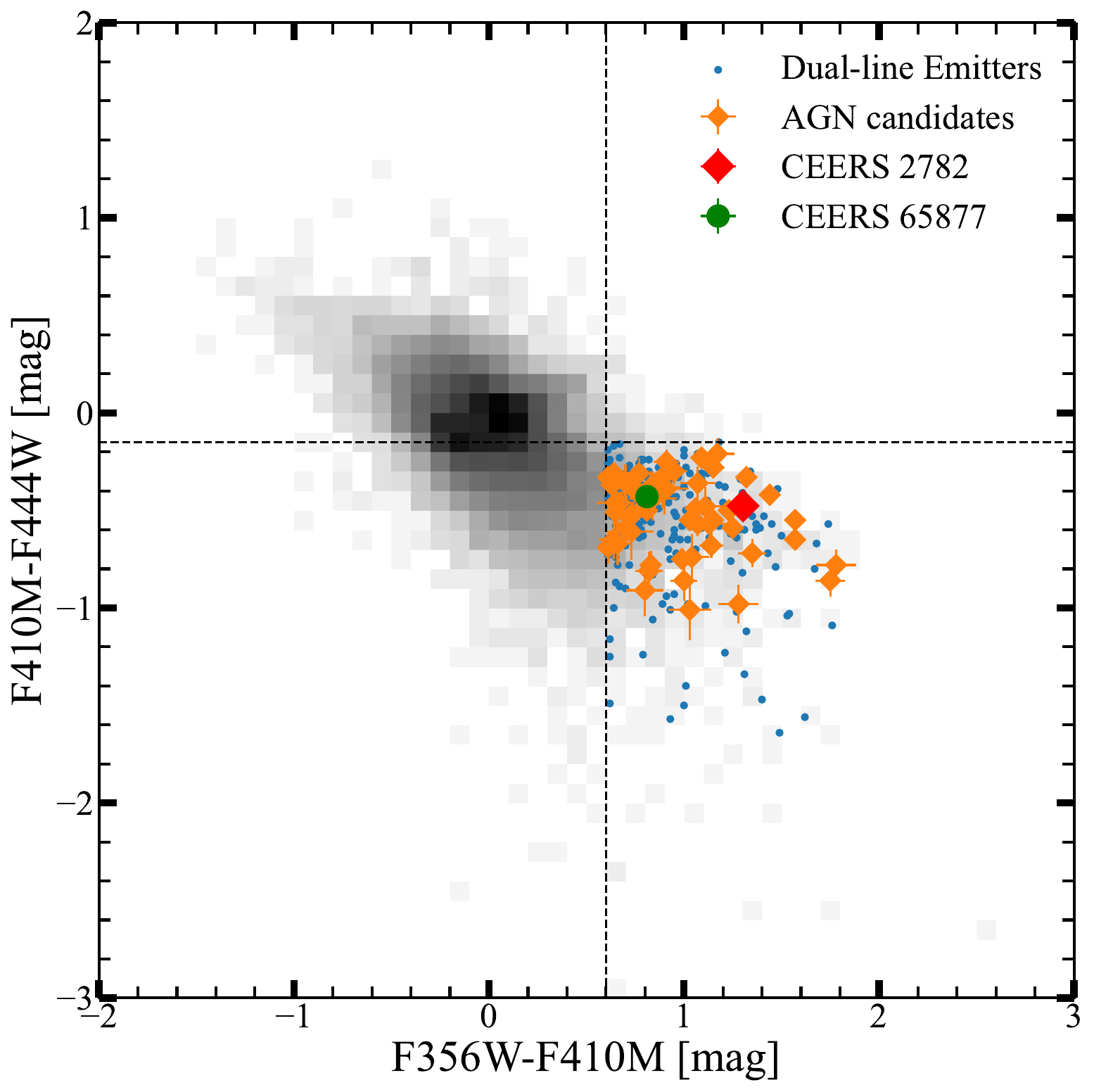}}\hspace{5mm}
{\includegraphics[width=85mm]{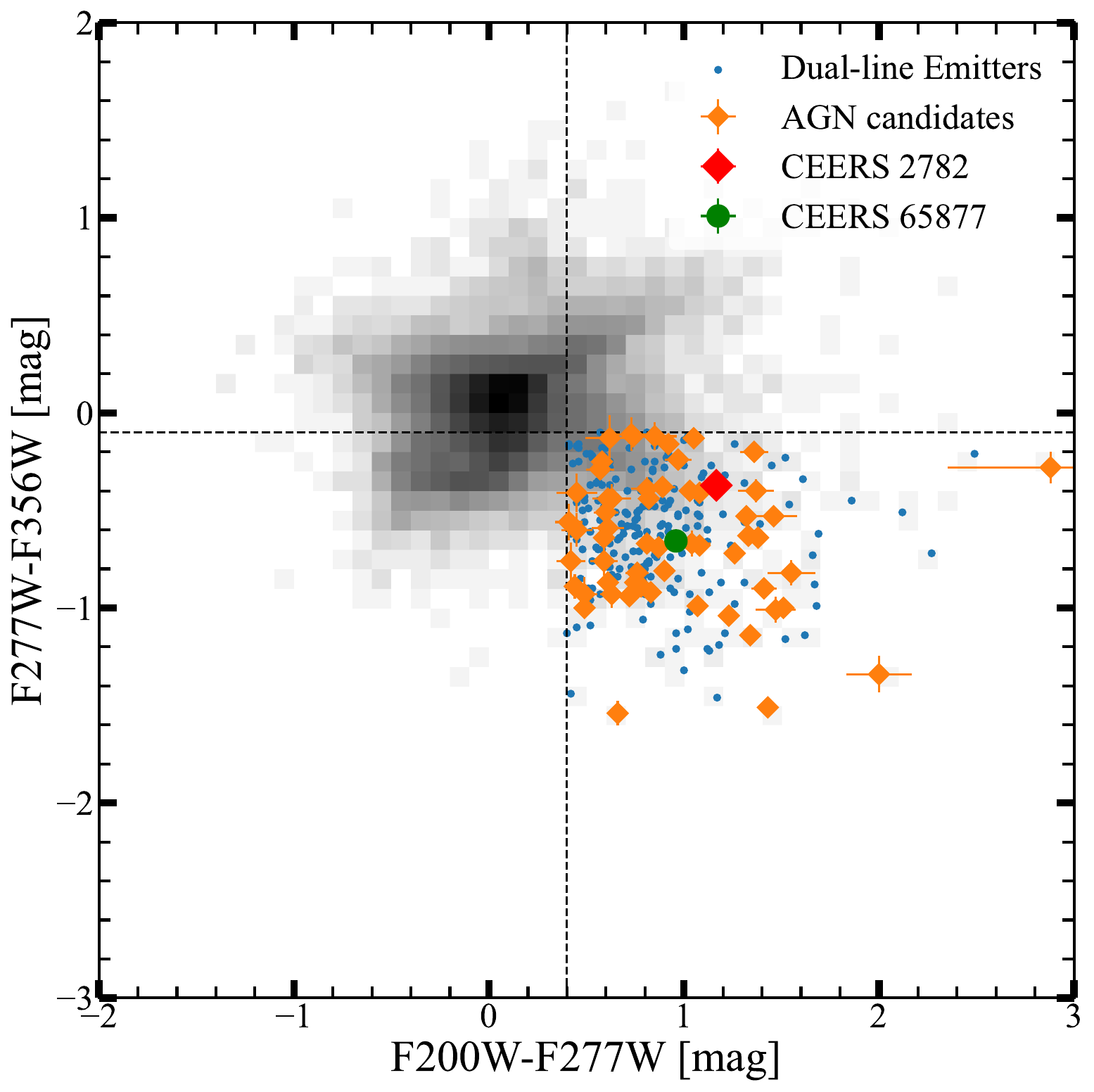}}	
\caption{The two-color plot of the sources in CEERS field. 
(\textit{Left:}) F410M $-$ F444W vs F356W $-$ F410M. 
(\textit{Right:}) F277W $-$ F356W vs F200W $-$ F277W. 
The dual-line emitters are shown as blue, while those with compact morphology (AGN candidates) are shown as orange.
The spectroscopically-confirmed objects, CEERS 2782 and CEERS 65877, are highlighted with red and green symbols, respectively.
The logarithmic density plot of the whole CEERS sources that satisfy the SNR thresholds (Step~2 in Table~\ref{table:scenario}) are shown with black.
}
\label{fig:fig_3_2color}      
\end{center}
\end{figure*}

%%%%%%%%
%	Fig.5.     %
%%%%%%%%
\begin{figure*}  
\begin{center}
{\includegraphics[width=85mm]{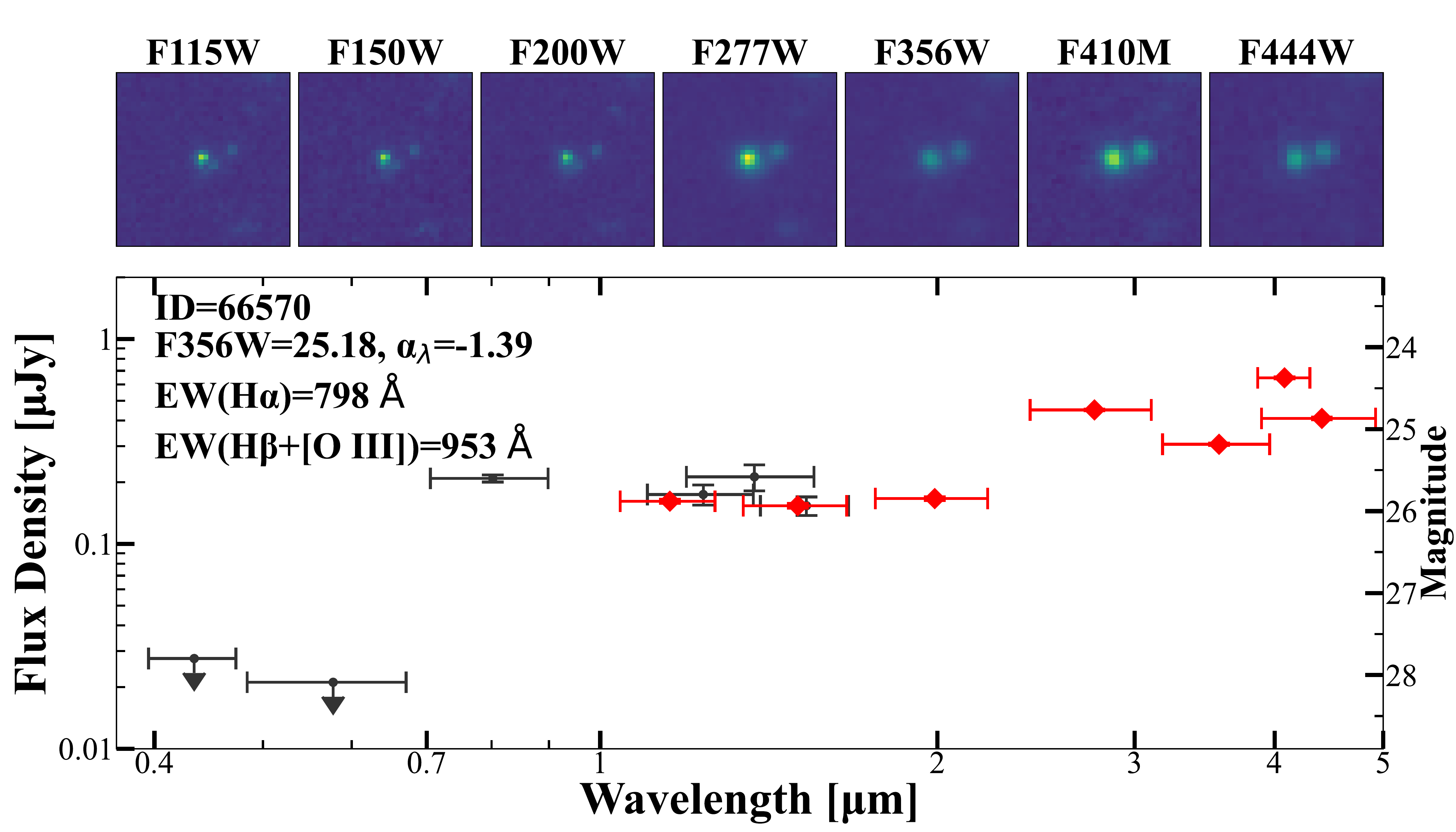}}\hspace{2mm}
{\includegraphics[width=85mm]{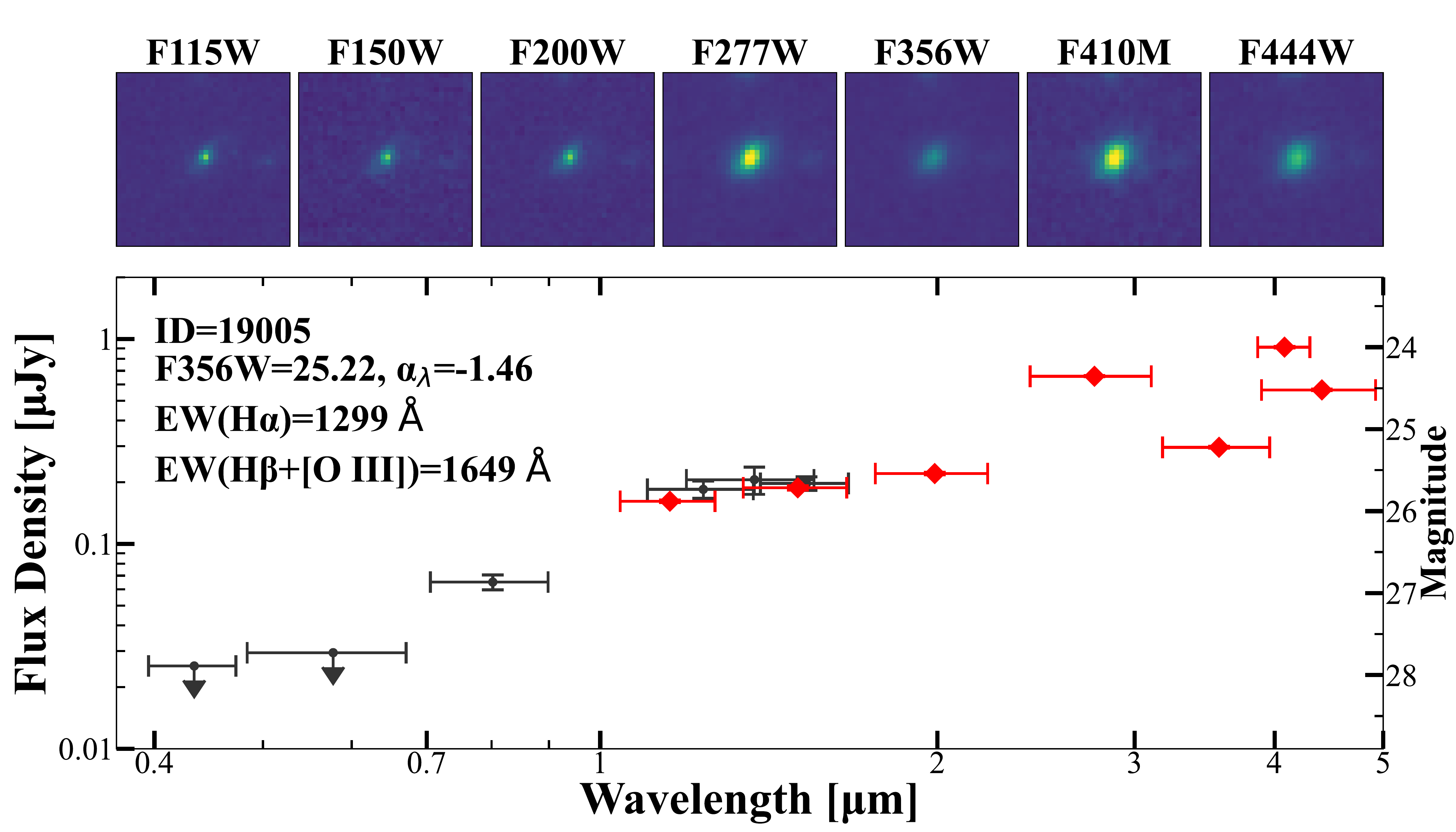}}	\\
{\includegraphics[width=85mm]{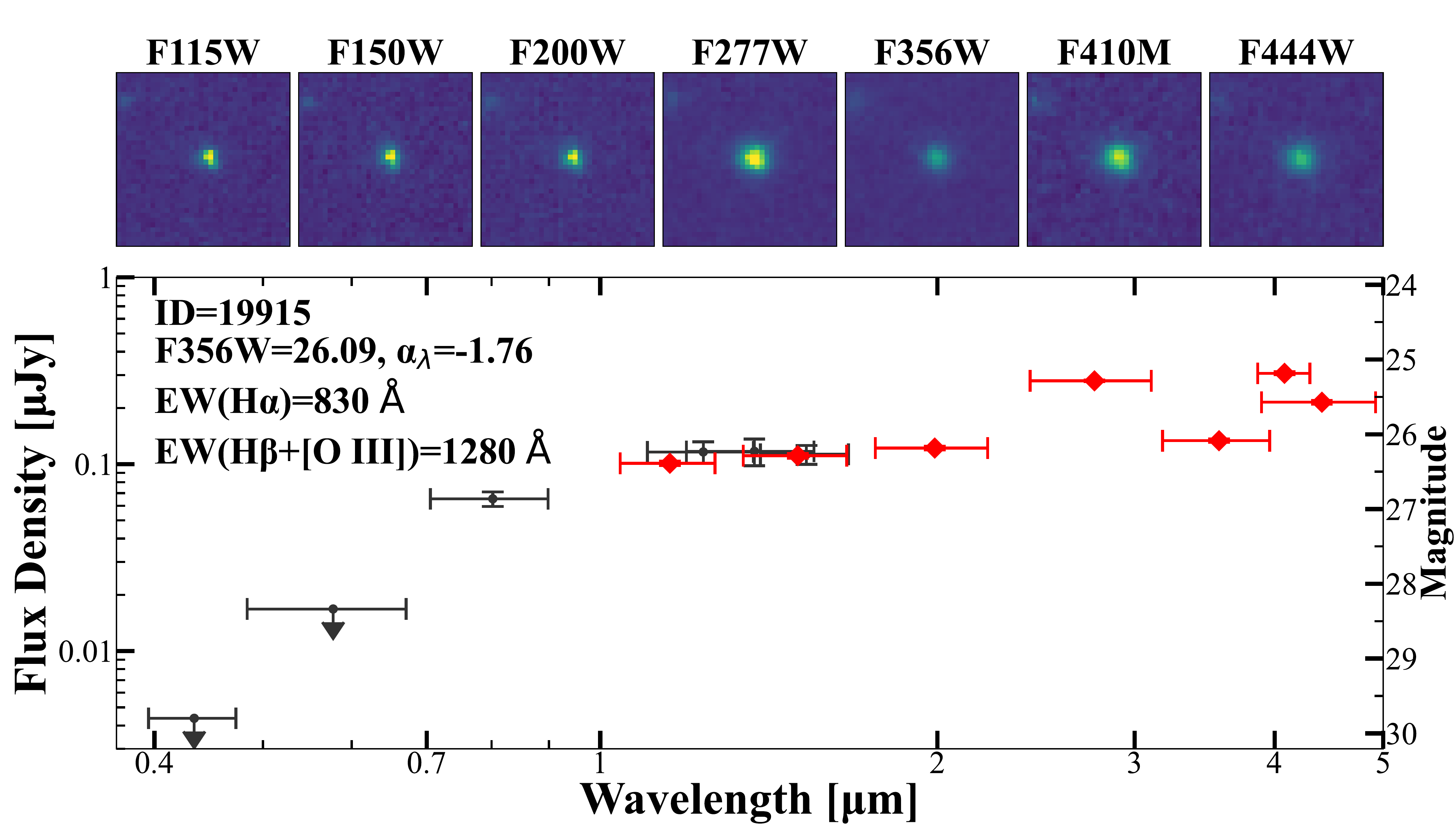}}\hspace{2mm}
{\includegraphics[width=85mm]{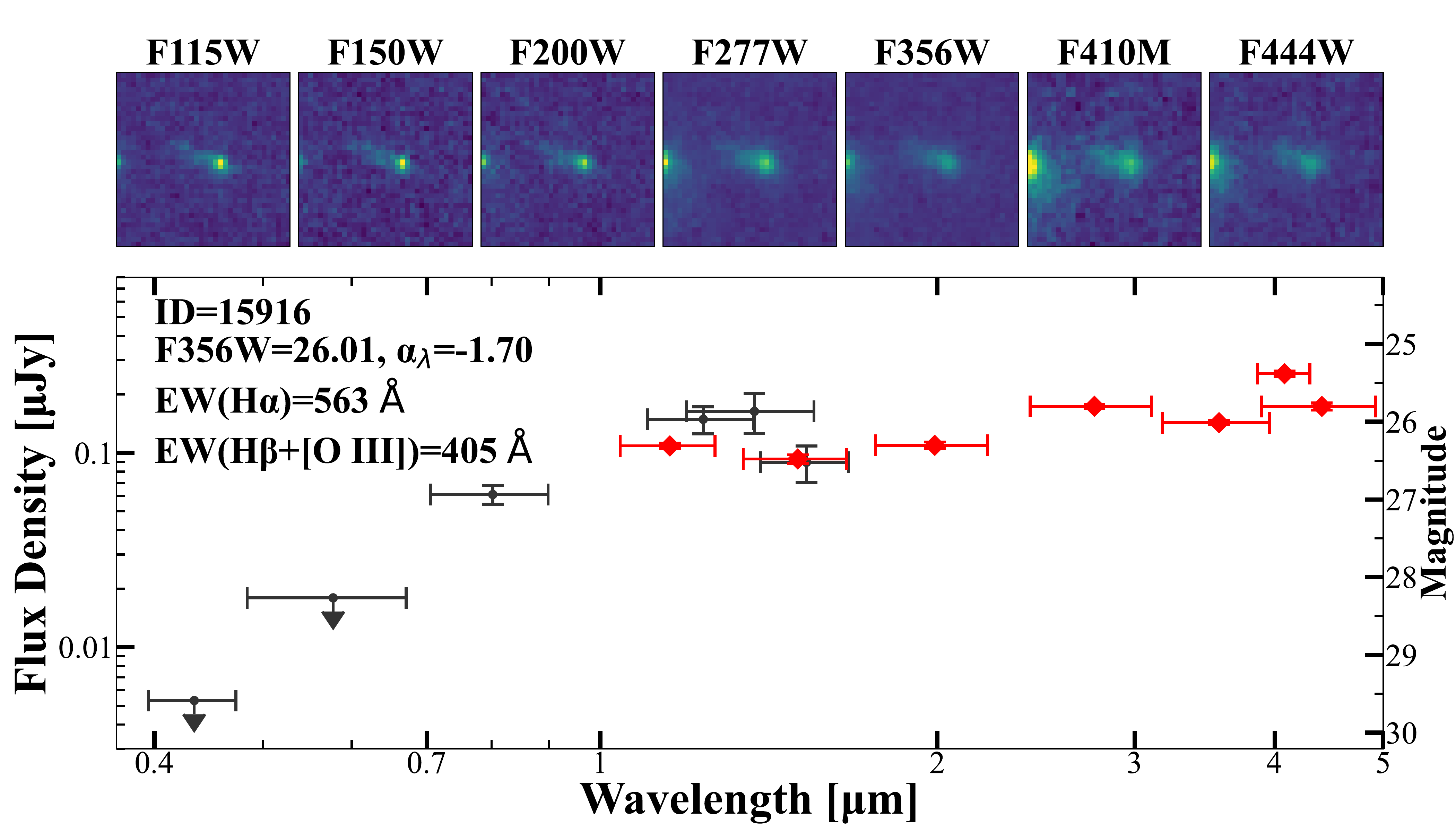}}	\\
{\includegraphics[width=85mm]{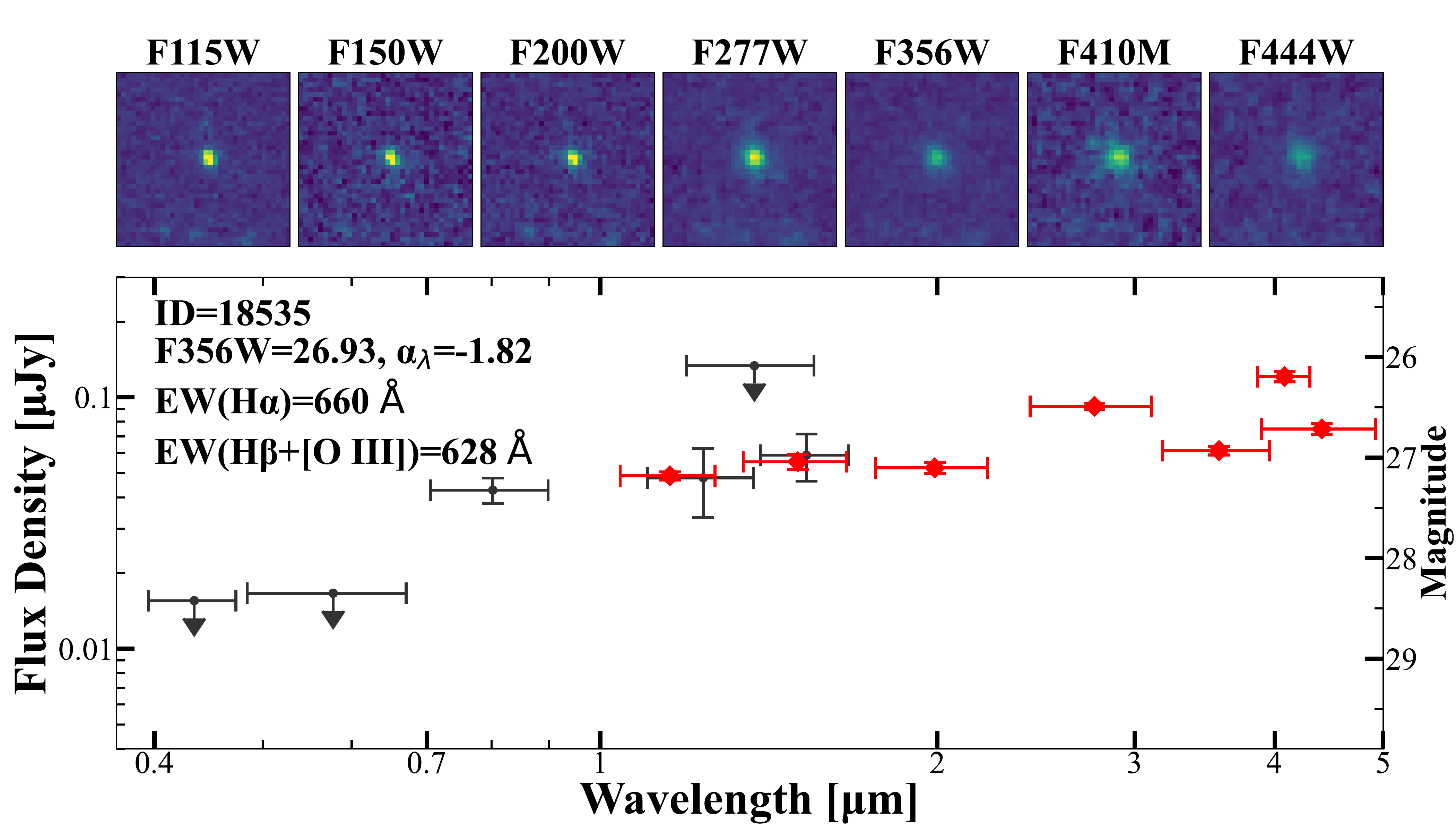}}\hspace{2mm}
{\includegraphics[width=85mm]{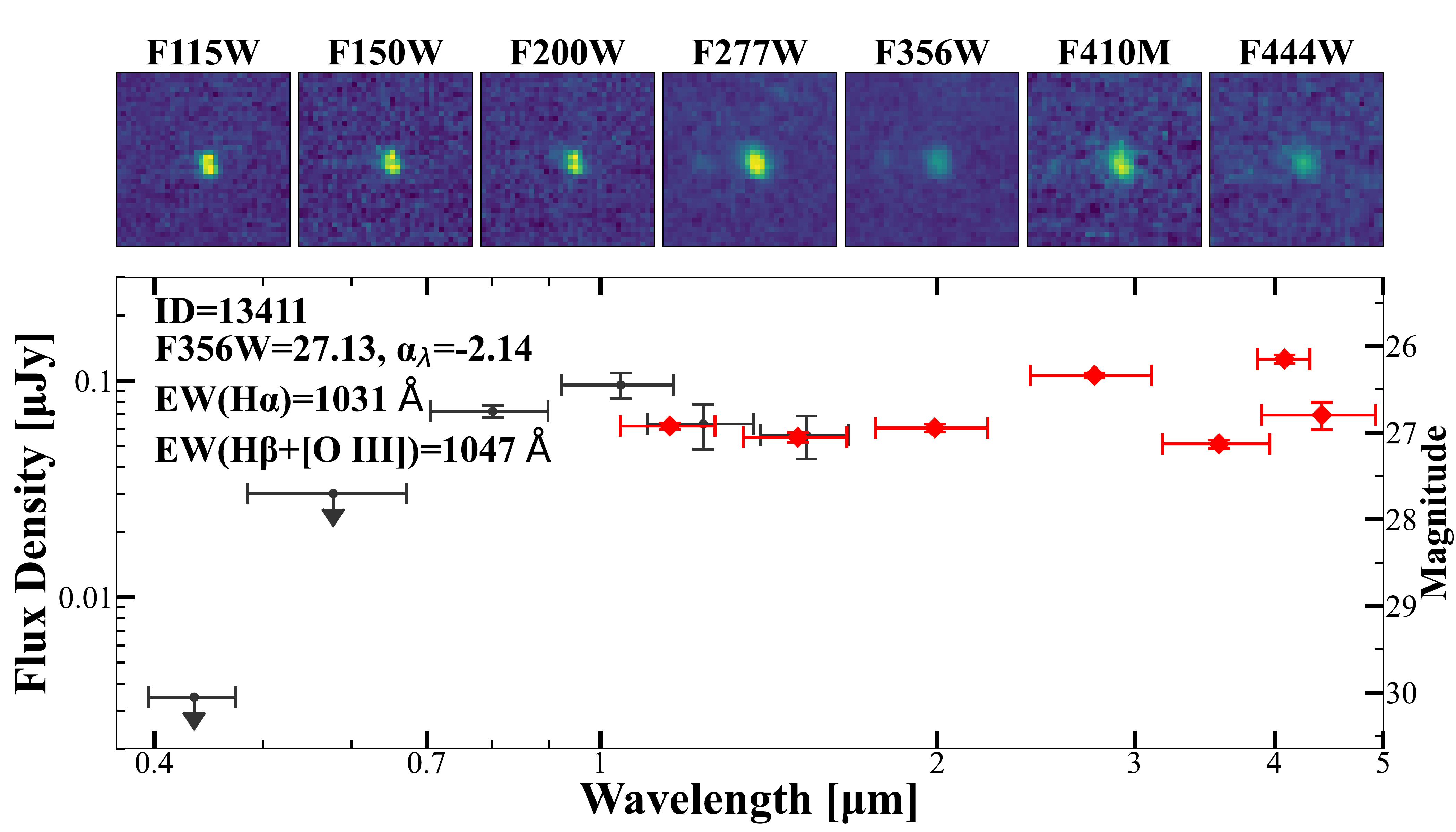}}	\\
{\includegraphics[width=85mm]{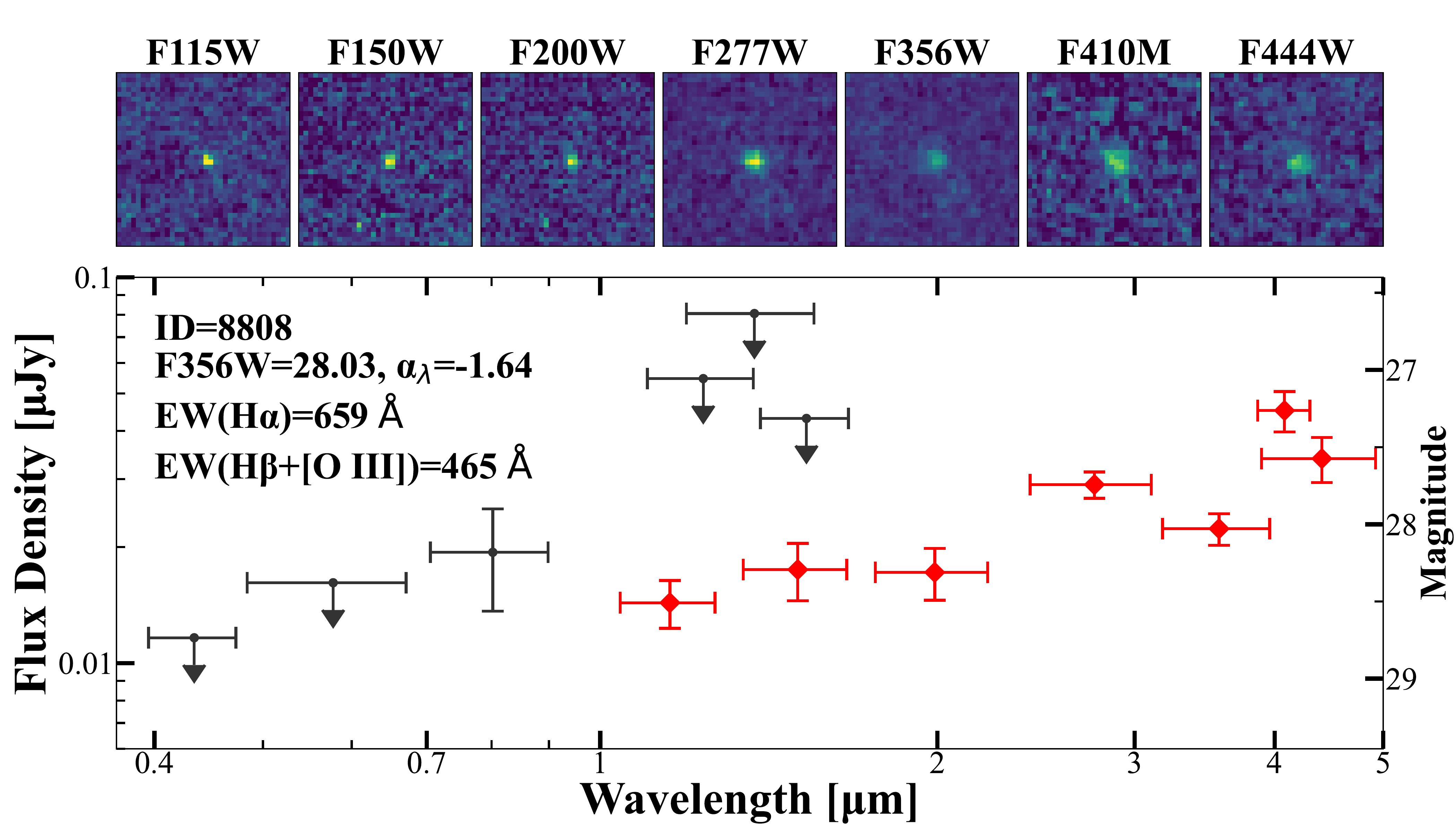}}\hspace{2mm}
{\includegraphics[width=85mm]{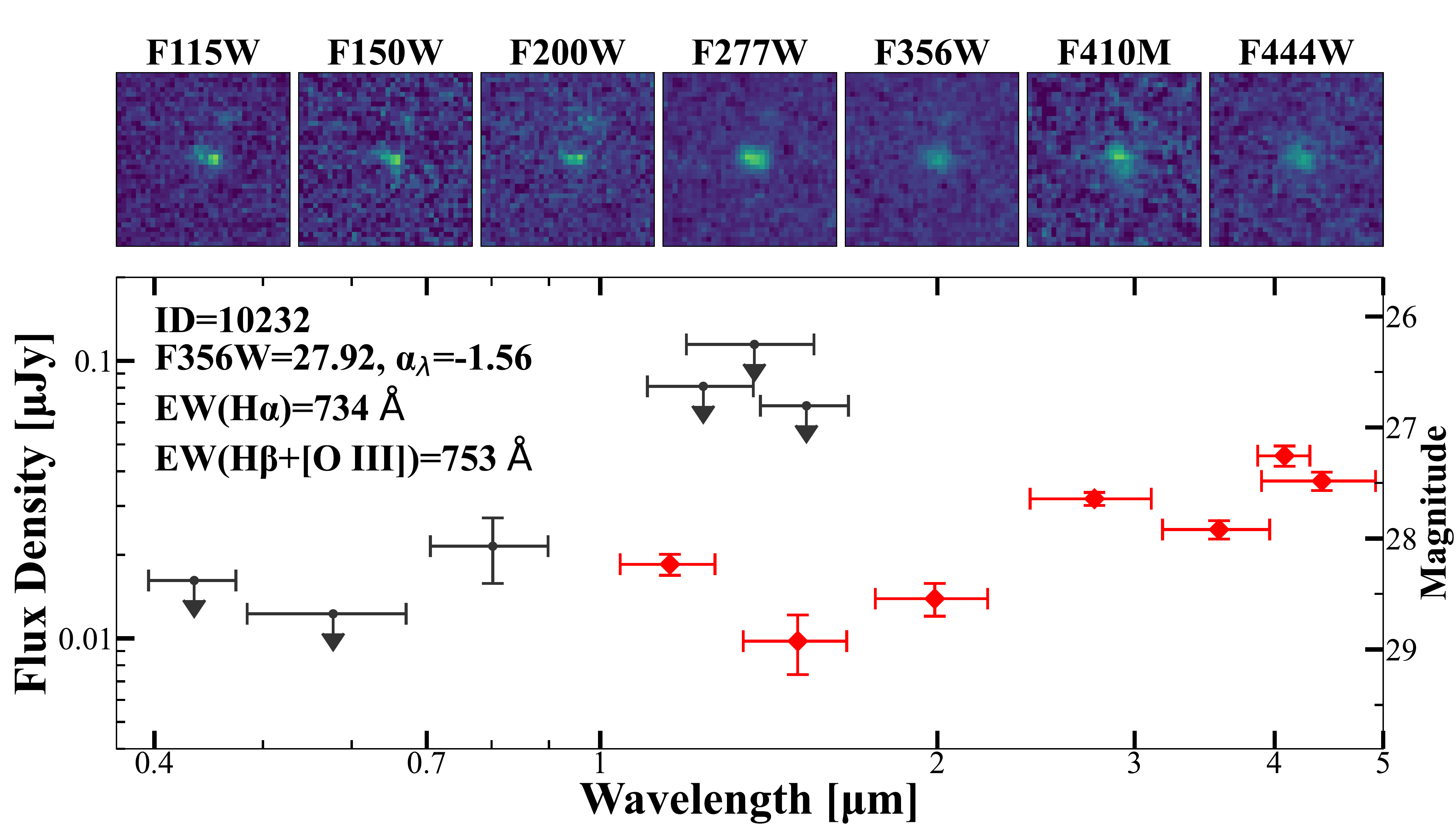}}	\\
\caption{
Eight examples of dual-line emitting galaxies from our sample, ordered by F356W magnitude with the brightest objects in the upper panels and the faintest in the lower panels.
For each object, we present the NIRCam images at the top and the rest-frame UV-to-optical SED at the bottom.
We show the F356W magnitude, continuum slope $\alpha_\lambda$, the rest-frame EWs of H$\alpha$ and \HbOIII\ in the legend of the SED panel.
The left four objects are among the AGN candidates with compact morphology, while the other four on the right represent the extended objects. 
The photometry data are taken from HST ACS/WFC3 (black) and JWST NIRCam (red). 
The $3\sigma$ upper limit flux is shown with arrows in the case of non-detection.
The SED plots of the remaining candidates can be found online (\url{https://github.com/JingsongGuo-astro/Dual\_Emitter}).
}
\label{fig:F356W25mag}            
\end{center}
\end{figure*}

\subsection{Compactness Measurement}\label{sec:CompMeas}

We measured the sizes of our dual-line emitter sample in order to identify compact objects, which we consider candidate AGN-dominated systems in this paper.
For this analysis, we used the size measurements derived from the DR 0.51 NIRCam images of the CEERS data \citep{Bagley_CEERS_2023}.
These measurements were based on two-dimensional surface brightness profile fitting carried out with the \textsc{Galfit} software version 3.0.5 \citep{Peng_2010_Galfit}.
The background-subtracted NIRCam images were used as inputs, with galaxy sizes measured independently in each of the six broadband images. 
To perform size measurements, bright star images were stacked across the entire CEERS fields for each broadband to generate empirical point spread function models. We  provide \textsc{Galfit} with noise images that account for both the intrinsic image noise (e.g., background noise and readout noise) as well as added Poisson noise due to the objects themselves. 
Each source was modeled with a single-S\'{e}rsic component to obtain the half-light radius $R_{\rm e}$.  All neighbouring objects down to three magnitudes fainter than the primary source in the fitting band are fit simultaneously using single S\'{e}rsic profiles. Sources fainter than this were masked during fitting.
We refer the readers to McGrath et al. (in prep.) for more details of this process.

We then applied the following additional criteria to extract compact sources:
\begin{align}
R_{\rm e, F115W}<0\arcsec.020\ \mathrm{or}\ 
R_{\rm e, F356W}<0\arcsec.058.
\end{align}
These thresholds correspond to the empirical image resolution for the two broadband filters\footnote{\url{https://jwst-docs.stsci.edu/jwst-near-infrared-camera/nircam-performance/nircam-point-spread-functions}}. 
We selected these two filters because they only cover the continuum emission of the $z=5$ galaxies, not affected by strong line emission.
%A total of 58 dual-line emitters  meet these criteria.
%In the subsequent sections of this paper, we regard these objects as candidates for AGN.

\section{Results} \label{sec:results}

\subsection{Dual-Line Emitters}\label{sec:DualEmitters}

By applying the selection methodology outlined in Table~\ref{table:scenario} (see also Section~\ref{sec:selection}) to our dataset, 
we identified 275 objects with significant color excesses both in the F410M and F277W bands. 
In Figure~\ref{fig:F356W25mag}, we highlight examples of four compact and four extended objects in the order of the F356W brightness,
to illustrate the diversity within our dual-line emitting galaxies and AGN samples.

It is worth emphasizing that our dual-line emitter search based on JWST photometric data does not employ the Lyman break 
selection criterion, which is commonly used for identifying high-redshift galaxies.
Nevertheless, most of our selected objects show photometric signatures indicative of a Lyman break, 
characterized by non-detections ($<3\sigma$) or significantly fainter magnitudes ($\geq 1$ mag) in the HST F435W and F606W bands 
compared to the HST F814W and JWST F115W bands.
We found that 15 out of all the 275 selected targets do not meet this Lyman-break expectation according to 
the HST photometry result in the DJA catalog. 
For these 15 objects, we visually checked the images of the HST F435W and F606W bands provided in 
the CANDELS and UVCANDELS projects \citep{Koekemoer_2011_candels}, and found that only one of them 
shows a counterpart in the HST images and thus  does not satisfies the Lyman break criterion.
Therefore, we decided to remove this one object 
%with clear detection in the HST images 
and keep the other 14 objects that have no detection in the HST photometry.
Additionally, we flagged 13 objects to be suspicious through a visual check of the NIRCam images, as they are close to the edge of the detector or have low surface brightness. 
We also removed them from our sample. 

Finally, we compile a catalog of 261 dual-line emitters, of which 58 (22\%) show compact morphology. 
In the subsequent sections of this paper, we regard these objects as unobscured AGN candidates.
%We classify those compact sources as AGN candidates.
Their distribution in the F356W $-$ F410M vs F410M $-$ F444W color plane and that for F200W $-$ F277W vs F277W $-$ F356W are presented in  Figure \ref{fig:fig_3_2color}.

\subsection{Continuum properties}\label{sec:ContProp}

For our 261 dual-line emitters (including extended galaxies and compact AGN candidates),
we model the continuum flux assuming a single power-law shape ($F_\lambda \propto \lambda^{\alpha_\lambda}$).
For this continuum fitting, we use the photometry at F115W, F150W, F200W, and F356W, because these four filters are not severely affected by  \HbOIII or  H$\alpha$ emission.
%This model utilizes four broadband photometry data in F115W, F150W, F200W, and F356W, where either of the prominent H$\alpha$ or \HbOIII emission lines does not enter.

%%%%%%%%
%	Fig.6.     %
%%%%%%%%
\begin{figure}  
\begin{center}
{\includegraphics[width=85mm]{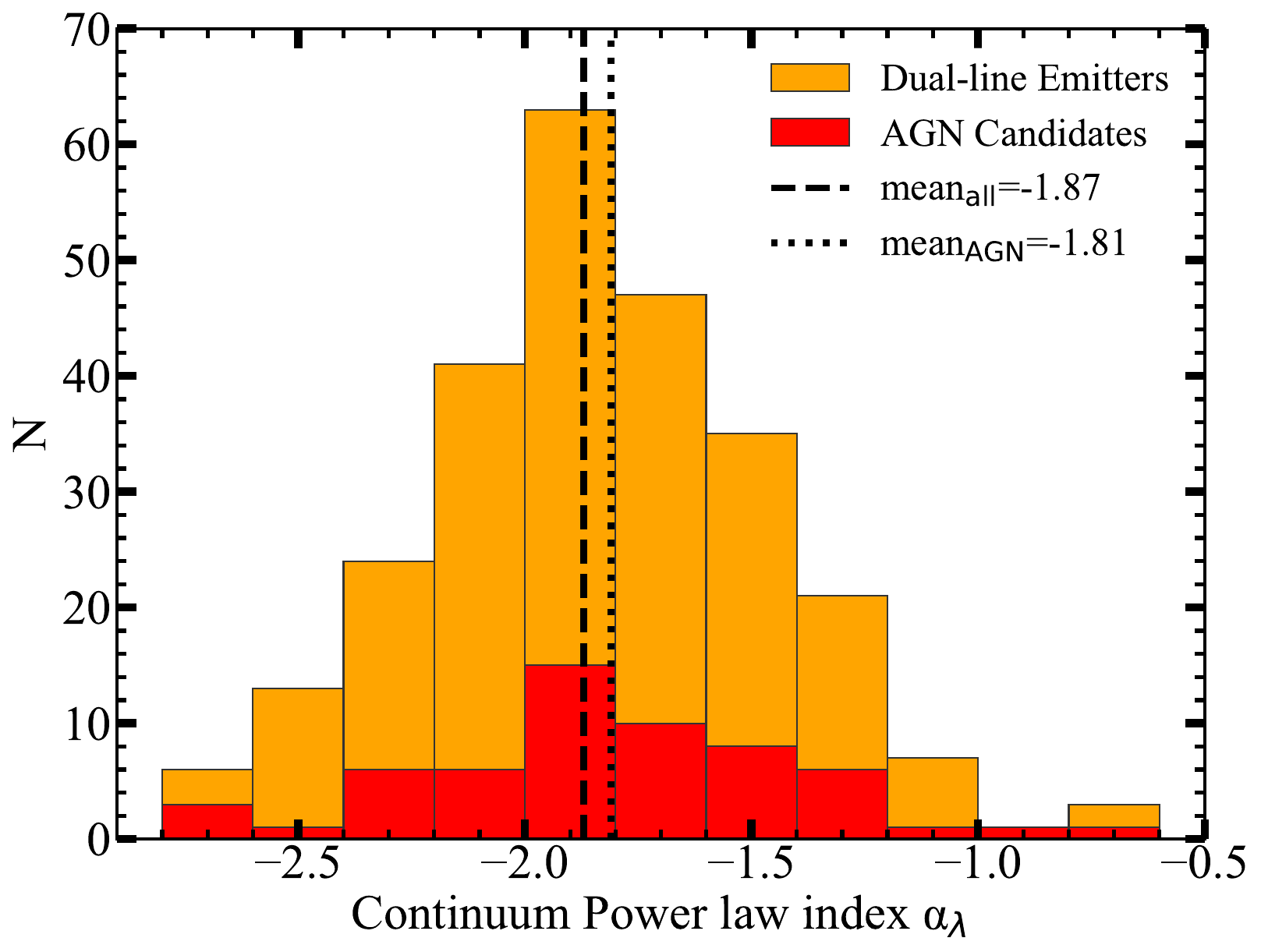}}	
\caption{The distribution of the continuum slope index $\alpha_\lambda$ of our dual-line emitters.
The whole sample is shown in orange, while the AGN candidates are shown in red. 
The mean value is  $\alpha_\lambda=-1.87$ (vertical dashed line) for the former with the standard deviation of 0.37, while $-1.81$ for the latter with the standard deviation of 0.40.}
%The mean value of all dual-line emitters is $\alpha_\lambda=-1.87$ (vertical dashed line) with standard deviation of 0.37. }
\label{fig:alHis}            
\end{center}
\end{figure}

Our sample shows an mean $\alpha_\lambda = -1.87$ with a standard deviation of 0.37. 
This value falls between the empirical value from AGN templates 
(e.g., $\alpha_\lambda \simeq -1.54$, \citealt{VandenBerk_2001}) 
and the median value of $z=5$ star-forming galaxies 
($\alpha_\lambda \simeq -2.3$ with median absolute UV magnitude $M_{\rm UV}^{\rm med}=-18.6$, \citealt{Topping_2024MNRAS}). 
Among our sample, 19 objects have an extreme blue continuum ($-2.8<\alpha_\lambda<-2.4$). 
These could be young metal-poor galaxies, as their UV continuum slope is consistent with 
previously detected Lyman-continuum emitting galaxies \citep{Chisholm_2022}.

Based on the continuum flux, we calculate the absolute UV magnitude ($\Muv$) at rest-frame 1450~\AA.
For all dual-line emitting galaxies, the UV magnitude ranges between $-23 < \Muv <-16$, while the most luminous AGN candidates reach $-20.5$ mag. 
The detection range for our AGN candidates aligns with other contemporary JWST research \citep[e.g.,][]{Maiolino_2023_JADES,Harikane_2023_agn},
but is $\sim 2$ magnitude fainter than the quasar population observed in ground-based surveys \citep{Niida_2020}. 
We discuss the UV luminosity function in Section~\ref{sec:UVLF}.

Using the continuum flux measurement for the AGN candidates, we calculate the bolometric luminosity based on 
the rest-frame 5100~\AA\ monochromatic luminosity $\lambda L_{5100}$, assuming that they are unobscured populations.
We adopt a bolometric correction factor of $L_{\rm bol}/\lambda L_{5100}=9.26$ \citep{Richards_2006} without correcting 
dust attenuation on $\lambda L_{5100}$.
The estimated bolometric luminosity is distributed at $10^{43}\lesssim L_{\rm bol}/({\rm erg~s}^{-1})\lesssim 10^{45}$. 
We further estimated the bolometric luminosity function in Section~\ref{sec:BLF}.

%%%%%%%%
%	Fig.7.     %
%%%%%%%%
\begin{figure*}  
\begin{center}
{\includegraphics[width=85mm]{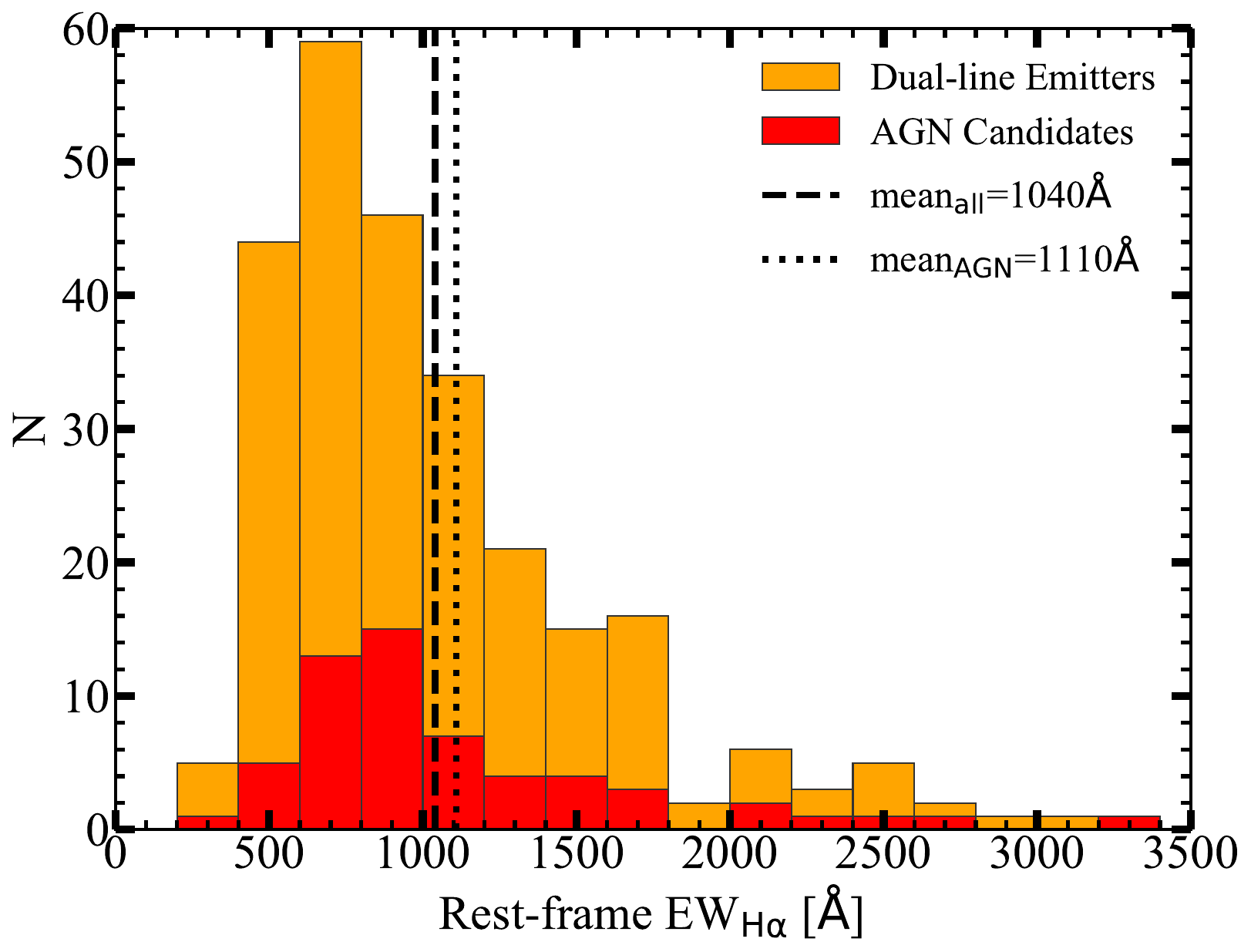}}\hspace{5mm}	
{\includegraphics[width=85mm]{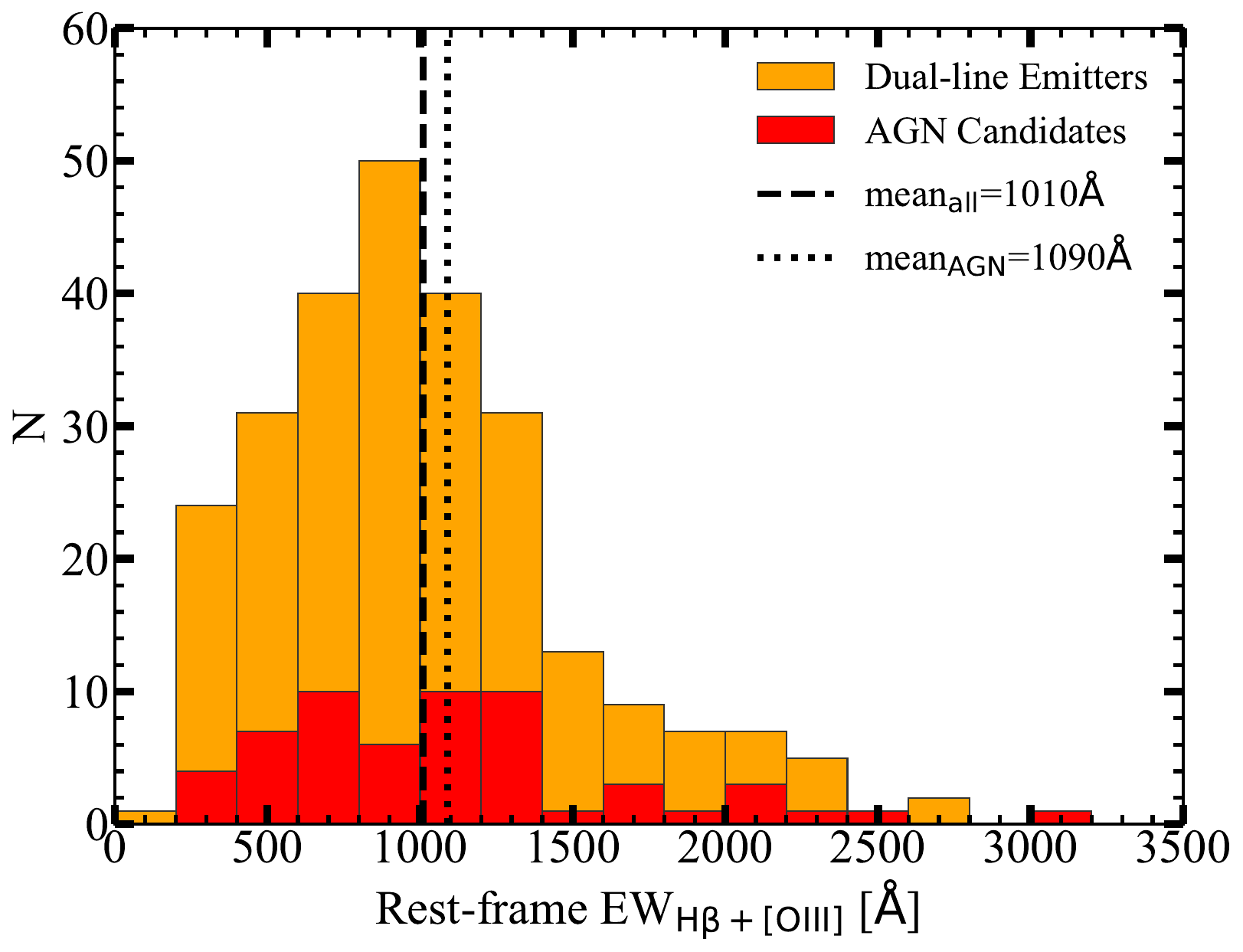}}	
\caption{The distribution of the rest-frame equivalent width of H$\alpha$ in F410M (Left) and \HbOIII\ in F277W (Right). 
All dual-line emitters are shown in orange, while the AGN candidates are shown in red. 
The mean values of all dual-line emitters are EW $= 1040$ \AA\ for H$\alpha$ and EW $= 1010$ \AA\ for \HbOIII, respectively (vertical dashed lines). For AGN candidates, the mean values are EW $= 1110$ \AA\ for H$\alpha$ and EW $= 1090$ \AA\ for \HbOIII, respectively }
\label{fig:EWHis}            
\end{center}
\end{figure*}

\subsection{Emission line properties} \label{sec:LineProp}

To quantify the color excess attributed to the \HbOIII and H$\alpha$ emission lines, we employ the F277W and F410M photometric flux
compared to the continuum flux level measured with the other four bands (see Section \ref{sec:ContProp}).
This approach enables us to calculate the observed equivalent widths for these emission lines, and convert these values to the rest-frame EW$_{\rm H\beta+ [O III]}$ and EW$_{\rm H\alpha}$ by assuming $z=5.2$, which is the median redshift within the range we select dual-line emitting galaxies. Other possible emission lines are neglected in this research.

In Figure \ref{fig:EWHis}, we present the distribution of the estimated rest-frame EW of H$\alpha$ and H$\beta+$[\ion{O}{3}].
Our sample of dual-line emitters has a mean EW$_{\mathrm{H\alpha}}=1040\ $\AA\ and EW$_{\mathrm{H\beta+[O III]}}=1010\ $\AA. 
The highest values of the two EWs are 3249~{\AA} and 3713~{\AA}, respectively.  
For H$\alpha$ emission, the distribution has good consistency with our selection threshold (EW$_{\rm H\alpha}\geq 500$~\AA). 
On the other hand, the distribution of EW$_{\rm H\beta+ [O III]}$ of our samples is extended to a value lower than 
the selection threshold of EW$_{\rm H\beta+ [O III]}\geq 650$~\AA).
The reason is that we might select dual-line emitters with weaker \HbOIII emission at the center of our redshift range (Section \ref{sec:selection}).

To further explore the contribution of the [\ion{O}{3}] luminosity from the \HbOIII emission complex, we use the ratio of 
$L_{\rm H\alpha}/L_{\rm H\beta}=3.1$ \citep{Osterbrock_1989} and separate the H$\beta$ contribution from the F277W flux.
The luminosity function of [\ion{O}{3}] emission together with that of H$\alpha$ emission are shown in Section~\ref{sec:LFO3Ha}.

%%%%%%%%
%	Fig.8.     %
%%%%%%%%
\begin{figure*}
\begin{center}
{\includegraphics[width=120mm]{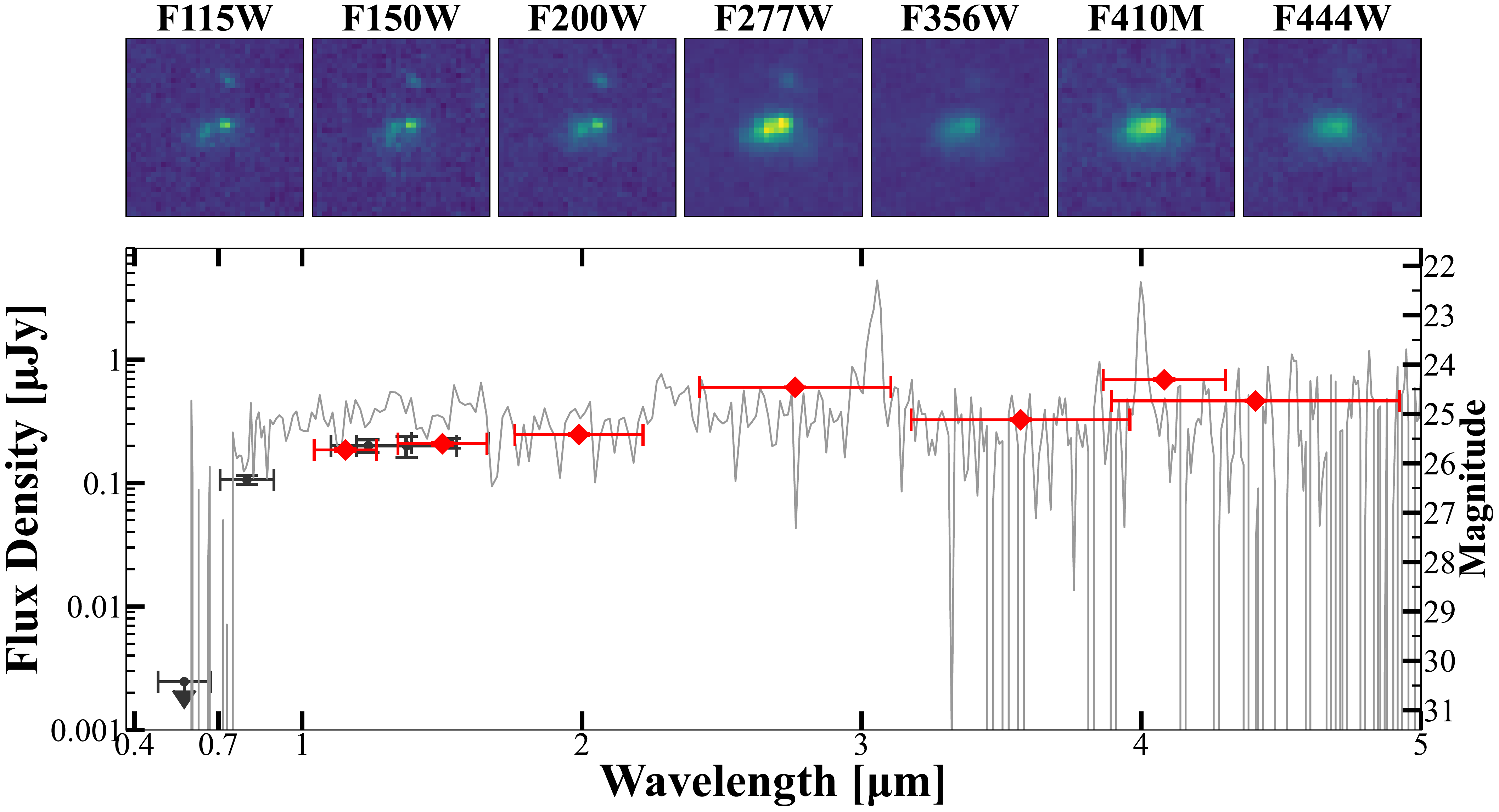}}
 \caption{A dual-line emitter with robust NIRSpec PRISM spectrum at $z=5.097$. The symbols are same to Figure \ref{fig:confirmedAGN} and \ref{fig:F356W25mag} 
 and the spectroscopy data is rescaled by a factor of 8 to fit the photometry data. 
This galaxy has been reported in \citet{Nakajima_2023}.}
\label{fig:prism_ex}            
\end{center}
\end{figure*}

\subsection{Recovery of known $z>5$ AGN} \label{sec:literature}

Previous studies have discovered a number of $z>4$ AGNs and candidates in the CEERS field, including modestly dust-reddened ones 
\citep{Kocevski_2023, Kocevski_2024, Harikane_2023_agn, Andika_2024, Kokorev_2024}.
Here we investigate overlaps of our dual-line emitters with those reported in the literature.

There are only a few spectroscopically-confirmed AGN at the redshift range of our interest ($z=5.03$--$5.26$).
Among the broad-line AGNs of \citet{Kocevski_2023} and \citet{Harikane_2023_agn}, CEERS~2782, the first confirmed JWST AGN at $z=5.24$ well satisfies the color criteria (Figure~\ref{fig:fig_3_2color}).
This is natural because our selection is based on this object. 
There is another object in \citet{Harikane_2023_agn}, CEERS~1465, that shows tentative detection of broad H$\alpha$ at $z=5.274$, which is at a slightly higher redshift than our sensitive range. It is not selected because its F277W$-$F356W ($=-0.03$ mag) and F356W$-$F410M ($=0.31$ mag) colors are weaker than our threshold.

We also check photometric samples of dust-reddened AGNs from \citet{Kokorev_2024} and \citet{Kocevski_2024}; however, we find no match among their $z\sim5$ AGN candidates. 
In the former study, 14 AGN candidates were reported with photometric redshifts of $4.5\leq z_{\rm phot}\leq 5.5$ while 12 candidates were reported in the latter. 
However, all of them have a positive or near-zero F277W$-$F356W color, which is indicative of a red continuum, but in conflict with our selection criteria.

\subsection{Existing spectroscopic data of our sample}

Here we present spectroscopic data available for the 261 dual-line emitters selected by this study.
In the DJA archive at the time of writing this paper, we find 4 objects observed by NIRSpec, including CEERS~2782 \citep[][Fig.~\ref{fig:confirmedAGN}]{Kocevski_2023}.

Out of the remaining three objects, one (CEERS~65877) shows robust spectra with visual inspection classified in the DJA archive and precise redshift measurements ($z_{\rm spec}=5.097$), 
with an extended morphology.
Figure~\ref{fig:prism_ex} presents the photometry and the PRISM spectrum of this object, showing prominent H$\alpha$ and \HbOIII emission lines
and the Lyman break feature.
The redshift value aligns with the redshift range for the color selection in the dual-line emitter search ($z_{\rm spec}=5.03-5.26$).
We further measure the rest-frame EW of H$\alpha$ and \HbOIII from spectroscopic data. We firstly take the median flux value from $2.5$ $\mu$m to $3.5$ $\mu$m and $3.5$ $\mu$m to $4.5$ $\mu$m as the continuum base of \HbOIII and H$\alpha$ lines, respectively.
Then we calculate the line flux by summing up the spectrum elements with $>1\mathrm{\sigma}$ detection above the continuum level.
Thus, the rest-frame EW of H$\alpha$ and \HbOIII are calculated to be $\mathrm{EW_{H\alpha}}=612\pm27~\mathrm{\AA}$ and 
$\mathrm{EW_{H\beta+[O III]}}=696\pm35~\mathrm{\AA}$. These EW values are consistent with our anticipation set in \ref{sec:Model}.
This PRISM spectrum further ensures the reliability of our selection criteria and the accuracy of our redshift estimations.

The currently available data is not sufficient to classify the other two objects.
One of them (ID~32653) was observed in the PRISM mode; however, the DJA spectrum misses data between 2.5 and 4.0 $\mu$m, in which \HbOIII\ and H$\alpha$ fall.
The PRISM spectrum of the other object (ID~43098) shows a noisy spectrum and we cannot confirm any emission lines.

\subsection{Little Red Dots} \label{sec:lrd}

We have also examined if our AGN candidates would be selected as LRD objects,
using NIRCam photometry selection criteria for LRDs at $z\gtrsim 4.5$ proposed by \citet{Greene_2024}:
\begin{align}
{\rm F115W}-{\rm F150W}&<0.8 \nonumber\\
\&~{\rm F200W}-{\rm F277W}&>0.7,\nonumber\\
\&~{\rm F200W}-{\rm F356W}&>1.0.
\label{eq:LRD}
\end{align}
With these selection conditions, none of our dual-line emitter with robust measurement of the F200W flux with
a ${\rm SNR}_{\mathrm{F200W}} > 5$ is classified as LRD.
This is because our selection requires a negative value of F277W $-$ F356W due to a color excess in F277W from the {\HbOIII} multiplet. 
However, most of the reported LRDs do not show emission lines strong enough to cause a color excess in the NIRCam broad and medium bands
(e.g. Section~4 in \citealt{Greene_2024}, Figure~2 in \citealt{Kocevski_2024}).
Thus, the selection method we have raised in this paper is not sensitive to LRDs.

Besides, though no object is classified as LRD, we find a number of objects with a blue UV continuum plus 
a red optical continuum (but not red enough to match the LRD criteria). For example, we show object 10232 
on the top left of Figure \ref{fig:F356W25mag}. 
It has a F115W$-$F150W $= -0.70$ and a F200W$-$F356W $= 0.62$ mag (0.38 mag smaller than the criterion) 
with EW$_{\rm H\alpha}=734$\ \AA\ and EW$_{\mathrm{H\beta+[O III]}}=753$\ \AA. 
These two broad-band colors might indicate a dust-reddened continuum. 
However, the reddening is moderate and cannot meet the LRD selection.
Thus, we consider that this method has the potential to select moderately dust-attenuated galaxies. 
This might help to match the gap between unobscured objects and LRDs.

\section{Luminosity function}\label{sec:LF}

We now investigate the statistical properties of the $z\sim5$ dual-line emitting galaxies identified in this work.
In this section, we show our measurements of the luminosity function (LF) as a function of 
UV magnitude ($\Muv$; Section~\ref{sec:UVLF}), bolometric luminosity ($L_{\rm bol}$; Section~\ref{sec:BLF}), H$\alpha$ and [\ion{O}{3}] luminosities 
($L_{\rm H\alpha}$ and $L_{\rm [OIII]}$; Section~\ref{sec:LFO3Ha}).
In the following sections, we discuss these LFs for the whole sample and for a sub-class of those that show compact morphology.

\begin{table*}
\renewcommand\thetable{2} %! fix indexing
\caption{Binned UV and bolometric luminosity function for our sample of AGN candidates and dual-line emitters}
\begin{center}
% \tablenum{1}
\begin{tabular}{ccccc}
\hline
\hline
&&UV LF &&\\
\hline
 &AGN candidates & &Dual-line emitters&\\
\hline
$\Muv$  & $N$ & $\log \Phi$& $N$ & $\log \Phi$ \\ 
 (mag) &  & ($\mpc^{-3}~{\rm mag}^{-1}$)& & ($\mpc^{-3}~{\rm mag}^{-1}$) \\ 
\hline

-23.0&&&1&$<-4.17$\\ 
-21.5&&&1&$<-4.17$\\
-21.0&&&2&$-4.17_{-0.53}^{+0.23}$\\
-20.5&1&$<-4.17$&9&-3.52$_{-0.18}^{+0.12}$\\
-20.0&3&-4.00$_{-0.37}^{+0.20}$&13&-3.36$_{-0.14}^{+0.11}$\\
-19.5&4&-3.87$_{-0.30}^{+0.18}$&16&-3.27$_{-0.12}^{+0.10}$\\
-19.0&12&-3.39$_{-0.15}^{+0.11}$&32&-2.97$_{-0.08}^{+0.07}$\\
-18.5&14&-3.33$_{-0.14}^{+0.10}$&55&-2.73$_{-0.06}^{+0.05}$\\
-18.0&12&-3.39$_{-0.15}^{+0.11}$&69&-2.63$_{-0.06}^{+0.05}$\\
-17.5&9&-3.52$_{-0.18}^{+0.12}$&51&-2.76$_{-0.07}^{+0.06}$\\
-17.0&3&-4.00$_{-0.37}^{+0.20}$&11&-3.43$_{-0.16}^{+0.11}$\\
-16&&&1&$<-4.17$\\

\hline
\hline
&&Bolometric LF &&\\
\hline
&AGN candidates&&&\\
\hline
$\log L_{\rm bol}$ & $N$ & $\log \Phi$ & &\\ 
(${\rm erg~s}^{-1}$) &  & ($\mpc^{-3}~{\rm dex}^{-1}$) & &\\
\hline 
43.25&3&-3.69$_{-0.37}^{+0.20}$&&\\
43.50&17&-2.94$_{-0.12}^{+0.09}$&&\\
43.75&16&-2.97$_{-0.12}^{+0.10}$&&\\
44.00&11&-3.13$_{-0.16}^{+0.12}$&&\\
44.25&5&-3.47$_{-0.26}^{+0.16}$&&\\
44.50&5&-3.47$_{-0.26}^{+0.16}$&&\\
44.75&1&$<-3.87$\\

\hline
\end{tabular}
\label{tab:LFdata1}
\end{center}
\tablecomments{~Column (1): UV magnitude (top) or bolometric luminosity (bottom). Column (2)\&(4): Number of objects in each bin. Column (3)\&(5): Number density. The errors reported are Poisson errors.
}
\vspace{5mm}
\end{table*}    

\begin{table*}
\renewcommand\thetable{3} %! fix indexing
\caption{Binned H$\alpha$ and [\ion{O}{3}] luminosity function}
\begin{center}
% \tablenum{1}
\begin{tabular}{ccccc}

\hline\hline
&&H$\alpha$ LF &&\\
\hline
 &AGN candidates&&Dual-line emitters&\\
\hline
$\log L_{\rm H\alpha}$ & $N$ & $\log \Phi$ & $N$ & $\log \Phi$\\ 
(${\rm erg~s}^{-1}$) &  & ($\mpc^{-3}~{\rm dex}^{-1}$) &  & ($\mpc^{-3}~{\rm dex}^{-1}$)\\
\hline
41.25&&&2&-3.87$_{-0.53}^{+0.23}$\\
41.50&10&-3.17$_{-0.17}^{+0.12}$&53&-2.45$_{-0.06}^{+0.06}$\\
41.75&18&-2.97$_{-0.12}^{+0.09}$&99&-2.18$_{-0.05}^{+0.04}$\\
42.00&16&-3.06$_{-0.12}^{+0.10}$&51&-2.46$_{-0.07}^{+0.06}$\\
42.25&3&-3.69$_{-0.37}^{+0.20}$&25&-2.77$_{-0.10}^{+0.08}$\\
42.50&7&-3.32$_{-0.21}^{+0.14}$&16&-2.97$_{-0.12}^{+0.10}$\\
42.75&2&-3.87$_{-0.53}^{+0.23}$&8&-3.27$_{-0.19}^{+0.13}$\\
43.00&2&-3.87$_{-0.53}^{+0.23}$&5&-3.47$_{-0.26}^{+0.16}$\\
43.50&&&2&-3.87$_{-0.53}^{+0.23}$\\
\hline\hline
&& [\ion{O}{3}] LF &&\\ \hline
  &AGN candidates&&Dual-line emitters&\\
\hline
$\log L_{[{\rm O III}]}$ & $N$ & $\log \Phi$ & $N$ & $\log \Phi$\\ 
(${\rm erg~s}^{-1}$) &  & ($\mpc^{-3}~{\rm dex}^{-1}$) &  & ($\mpc^{-3}~{\rm dex}^{-1}$)\\
\hline
41.00&1&$<-3.87$&4&-3.57$_{-0.30}^{+0.18}$\\
41.25&1&$<-3.87$&6&-3.40$_{-0.23}^{+0.15}$\\
41.50&4&-3.57$_{-0.30}^{+0.18}$&32&-2.67$_{-0.08}^{+0.07}$\\
41.75&5&-3.47$_{-0.26}^{+0.16}$&31&-2.68$_{-0.09}^{+0.07}$\\
42.00&14&-3.03$_{-0.14}^{+0.10}$&55&-2.43$_{-0.06}^{+0.05}$\\
42.25&12&-3.09$_{-0.15}^{+0.11}$&54&-2.44$_{-0.06}^{+0.06}$\\
42.50&10&-3.17$_{-0.17}^{+0.12}$&36&-2.61$_{-0.08}^{+0.07}$\\
42.75&5&-3.47$_{-0.26}^{+0.16}$&20&-2.87$_{-0.11}^{+0.09}$\\
43.00&5&-3.47$_{-0.26}^{+0.16}$&11&-3.13$_{-0.16}^{+0.11}$\\
43.25&1&$<-3.87$&5&-3.47$_{-0.26}^{+0.16}$\\
43.50&&&2&$-3.87_{-0.53}^{+0.23}$\\
43.75&&&1&$<-3.87$\\
44.00&&&1&$<-3.87$\\
\hline
\end{tabular}
\label{tab:LFdata2}
\end{center}
\vspace{5mm}
\end{table*}

%%%%%%%%
%	Fig.9.     %
%%%%%%%%
\begin{figure*}
\begin{center}
{\includegraphics[width=130mm]{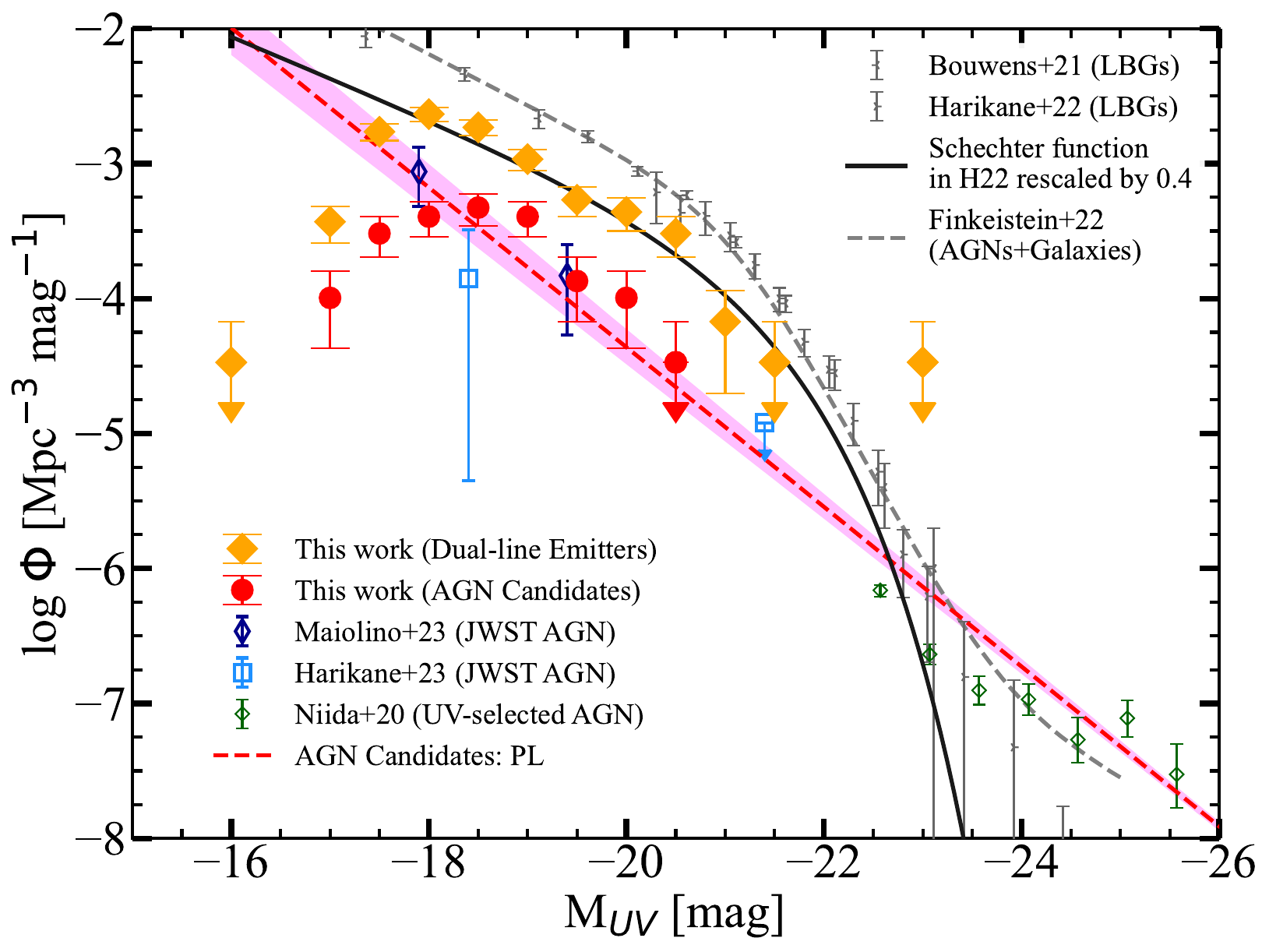}}
\caption{The $z\sim 5$ UV luminosity function of dual-line emitting galaxies (orange) and unobscured AGNs (red) without completeness correction.
The observed abundance of our AGN candidates is higher than the quasar abundance from ground-based surveys by more than two orders of magnitude \citep[green symbols, ][]{Niida_2020},
but in contrast aligns with that of unobscured AGNs at the fainter regime of $\Muv>-20$ reported through JWST surveys: 
\citet{Harikane_2023_agn} (cyan) and \citet{Maiolino_2023_JADES} (blue).
The LF of Lyman-break galaxies at $z\sim 5$ are presented with grey symbols and curve \citep{Bouwens_2021,Harikane_2022a,Finkelstein_Bagley_2022}, 
and the LF fitting with a Schechter function scaled by a factor of 0.4 (solid black curve) agrees with that of our dual-line emitting galaxies, 
indicating that our dual-line emitter selection efficiently identifies $\sim 40\%$ of the total star-forming galaxy population in the redshift range. For AGN candidates, we fit the AGN LF with a power-law function (dashed curve), combining both AGN candidates in this work and ground-based selected AGNs in \citet{Niida_2020}. The result shows that the slope at the low brightness end is steeper than the previous estimation of the ground-base surveys.
}
\label{fig:UVLF}            
\end{center}
\vspace{2mm}
\end{figure*}

To assess the abundance of dual-line emitting galaxies and unobscured AGNs in the CEERS field, we calculate the comoving volume 
over the redshift range $z=5.03$--$5.26$,
\begin{equation}
V_{\rm c} = \frac{c\Delta \Omega}{H_0}\int^{z_{\rm max}}_{z_{\rm min}}\frac{D_{\rm L}^2(z)}{E(z)(1+z)^2}dz,
\end{equation}
where $D_{\rm L}(z)$ is the luminosity distance and $E(z)=[\Omega_{\rm m}(1+z)^3 + \Omega_\Lambda]^{1/2}$.
The minimum and maximum redshift are set to $z_{\rm min}=5.03$ and $z_{\rm max}=5.26$, respectively. 
The effective survey coverage of CEERS is $\Delta \Omega = 97.0~{\rm arcmin}^2$ \citep{Bagley_2023}.
Thus, the survey volume of this study is estimated to be $59,340\ \mathrm{cMpc^3}$.
Note that selection completeness is not corrected for in this work.
Tables~\ref{tab:LFdata1} and \ref{tab:LFdata2} summarizes the binned LF data  for the whole sample and that for the compact dual-line emitters.
We describe more details for each of these in the following sections.

%%%%%%%%
%	Table 2    %
%%%%%%%%
\begin{table}[t]
\renewcommand\thetable{4} %! fix indexing
\caption{Best-fit parameters of the UV, bolometric, H$\alpha$, and [\ion{O}{3}] luminosity functions.}
\begin{center}
% \tablenum{1}
\begin{tabular}{cc}
\hline
\hline
UV luminosity & DPL \\
\hline
$\Phi_{\rm UV}^*$ [cMpc$^{-3}$ mag$^{-1}$]& $1.0^{+0.8}_{-0.5}\times10^{-9}$ \\
$\alpha$ & $-2.41 \pm 0.06$ \\
$\beta$  & $-4.9\pm 1.2$ \\
$M_{\rm UV}^*$ & $-28.0\pm0.3$ \\
\hline
\hline
Bolometric luminosity & DPL \\
\hline
$\Phi_{\rm bol}^*$ [cMpc$^{-3}$ dex$^{-1}$]& $3.7^{+8.3}_{-2.5}\times10^{-4}$ \\
$\bar{\alpha}$ & $0.62 \pm 0.12$ \\
$\bar{\beta}$  & $1.72 \pm 0.12$ \\
$L_{\rm bol}^{*}$ [erg s$^{-1}$]& $3.2^{+5.9}_{-2.1}\times10^{44}$ \\
\hline
\hline
H$\alpha$ luminosity & PL \\
\hline
$\Phi_{\rm H\alpha}^*$ [cMpc$^{-3}$ dex$^{-1}$] & $6.0^{+2.2}_{-1.6}\times10^{-4}$ \\
$\bar{\alpha}$ & $1.39 \pm 0.23$ \\
$L_{\rm H\alpha}^{*}$ [erg s$^{-1}$]& $10^{42}$(fixed) \\
\hline
\hline
[\ion{O}{3}] luminosity & PL \\
\hline
$\Phi_{[{\rm OIII}]}^*$ [cMpc$^{-3}$ dex$^{-1}$]& $1.02^{+0.21}_{-0.17}\times10^{-3}$ \\
$\bar{\alpha}$ & $0.50\pm0.16$ \\
$L_{\rm [OIII]}^{*}$ [erg s$^{-1}$]& $10^{42}$(fixed) \\
\hline
\end{tabular}
\label{tab:fit}
\end{center}
\tablecomments{~The UV and bolometric luminosity functions are fitted with a DPL form, while the H$\alpha$ and [\ion{O}{3}] 
luminosity functions are fitted with a single power-law form.
}
\vspace{5mm}
\end{table}

%%%%%%%%
%	Fig.10.     %
%%%%%%%%
\begin{figure*}
\begin{center}
{\includegraphics[width=130mm]{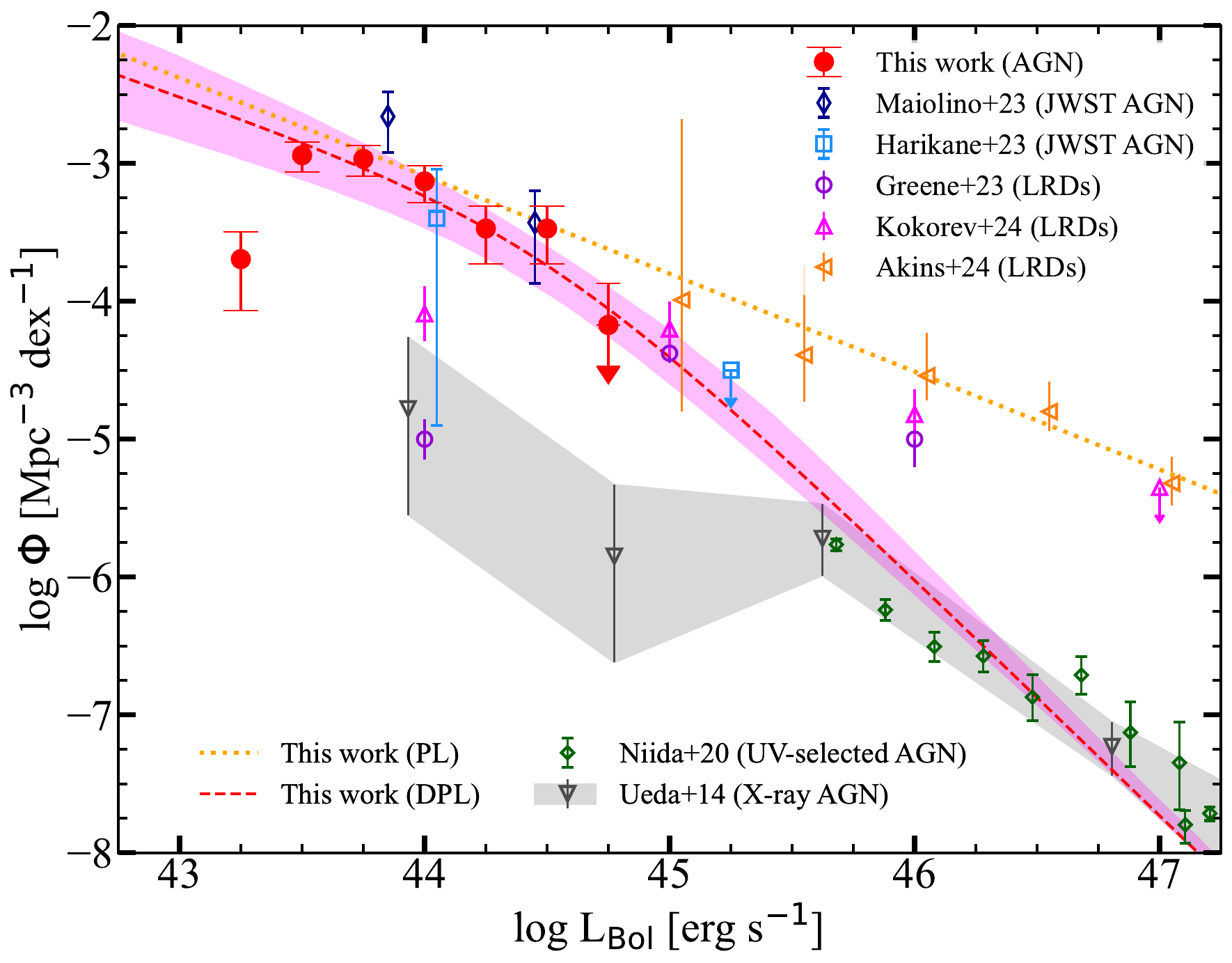}}	
\caption{AGN bolometric luminosity function at $z\sim5$ and the symbols are same to Figure \ref{fig:UVLF}. 
For comparison, the bolometric luminosity function data for dust-reddened AGNs (LRDs) are presented 
(\citealt{Greene_2024}, purple; \citealt{Kokorev_2024}, magenta;
\citealt{Akins_2024LRD}, orange). 
For our candidates, the bolometric luminosity is calculated from the rest-optical $\lambda L_{5100}$ luminosity. The results of ground-based survey is also presented \citep{Niida_2020}, which is converted from UV magnitude. The grey inverted triangles and shaded region present the results from X-ray detection \citep{Ueda_2014}, which is converted from X-ray luminosity. The grey dotted curve presents the DPL fitting connecting our AGN candidates and ground base survey and the shaded region represents the 1$\sigma$ confidence interval. The grey dashed line shows the power law fitting of our candidates and LRDs.
}
\label{fig:Lbol}
\end{center}
\vspace{2mm}
\end{figure*}

\subsection{UV luminosity function}\label{sec:UVLF}

Figure~\ref{fig:UVLF} illustrates the binned UV luminosity functions of both the whole sample (orange) and the unobscured AGN candidates (red).
We set a magnitude bin size to $\Delta \Muv=0.5$ mag from $\Muv=-23$ to $-16$ mag.
We find that the UV LF of the dual-line emitters has a similar shape to that for $z\sim5$ Lyman break galaxies (LBGs) at UV magnitude down to $\Muv = -18$ mag \citep{Bouwens_2021, Harikane_2022a, Finkelstein_Bagley_2022}.
In Figure~\ref{fig:UVLF}, we show that the binned UV LF is well consistent with the best-fit Schechter function of \citet{Harikane_2022a} after rescaling by a factor of 0.4. We emphasize that this factor should be treated as a lower limit, since we have not corrected any detection incompleteness.
We observe a decline of the number density at $\Muv>-18$ both in the whole sample and the compact sub-sample.
This is likely affected by selection incompleteness at the faint end.
Dual-line emitters in this study thus represent a substantial fraction of the UV-selected galaxy population.

Our selection finds that $22\%$ of the dual-line emitting galaxies qualify as unobscured AGN candidates (Sec.~\ref{sec:DualEmitters}), which corresponds to $(10\pm1)\%$ of LBGs observed at the same redshift range, applying the fraction of the dual-line emitter to LBG, 40\%.
This fraction is consistent with the AGN fraction ($f_{\rm AGN}\sim 5-10\%$) reported by previous JWST results that cover similar magnitude ranges 
\citep[e.g.,][]{Maiolino_2023_JADES, Harikane_2023_agn}.
We emphasize that it is intriguing to find that our selection  purely based on JWST data finds a similarly high fraction of AGN (candidates), because these early JWST spectroscopic observations preferentially targeted UV-bright galaxies or have complex selection functions.

The observed abundance of AGN candidates is estimated as $\Phi \simeq 3\times 10^{-4}~\mpc^{-3}~{\rm mag}^{-1}$ at $\Muv \simeq -18$ mag.
This value is more than two orders of magnitude higher than the extrapolation of the quasar LF obtained by ground-based studies
\citep[see e.g.,][]{Niida_2020}.
This significant discrepancy has been consistently reported for various types of AGNs, 
both unobscured and dust-reddened populations identified in JWST surveys 
\citep[e.g.,][]{Onoue_2023,Kocevski_2023,Harikane_2023_agn,Matthee_2024,Greene_2024,Kocevski_2024,Kokorev_2024}.
The abundance of unobscured AGN candidates in our sample closely matches those reported by 
\citet[][blue diamond]{Maiolino_2023_JADES} and \citet[][cyan square]{Harikane_2023_agn}.

The brightest population in our AGN candidates reach $\Muv \simeq -20.5$ mag, which is still fainter than the quasars observed in the ground-based surveys.
Although we cannot directly compare our LF with the ground-based surveys at a matched magnitude range, we here model the UV LF over a wide magnitude range using the compact dual-line emitters from this study and brighter quasars from \citet{Niida_2020}.
We use a double power-law (DPL) function to fit the two binned LF measurements.
The DPL function is described as
\begin{equation}
\Phi(\Muv)=
\dfrac{\Phi^\ast_{\rm UV}}{10^{0.4(\alpha+1)\Delta \Muv}+10^{0.4(\beta+1)\Delta \Muv}}
\end{equation}
where $\Phi^\ast_{\rm UV}$ (in units of Mpc$^{-3}$ mag$^{-1}$) is the overall normalization, $\Delta \Muv = \Muv-\Muv^\ast$,
$\Muv^\ast$ is the characteristic magnitude, $\alpha$ and $\beta$ are the faint- and bright-end slopes, respectively.
We here incorporate the data from both our AGN candidates (red) and unobscured AGNs cataloged in ground-based surveys 
\citep[green,][]{Niida_2020}, focusing on AGNs with a UV magnitude brighter than
$\Muv<-18$ in the fitting to ease the effects of selection incompleteness.
The best-fit parameters are listed in Table~\ref{tab:fit}.

We note that the characteristic UV magnitude is as bright as $\Muv^{\star}= -28.0\pm0.3$ mag,
so that the LF data shown in Figure~\ref{fig:UVLF} is effectively fitted by a single power-law function.
The fitted faint-end slope of $\alpha=-2.41^{+0.06}_{-0.06}$ is relatively steeper than $\alpha=-2.00^{+0.40}_{-0.03}$ 
estimated by \citet{Niida_2020}, due to the high abundance of our AGN candidates within the CEERS field. 
For the bright-end slope, our fitted result of $\beta=4.8\pm1.2$ agrees with \citet{Niida_2020} in the margin of uncertainty.
We also test a single power-law fitting in the range of $\Muv= -18$ to $-27.9$ mag. The resulting slope is $-2.48\pm0.05$, which is in agreement with the faint-end slope suggested by DPL fitting.

\subsection{Bolometric AGN luminosity function} \label{sec:BLF}

Here we only consider the compact dual-line emitters to discuss the bolometric AGN luminosity function.
Figure~\ref{fig:Lbol} shows the bolometric luminosity function for our unobscured AGN candidates at $z\sim 5$
and a compilation of multi-wavelength studies: 
rest-UV selected unobscured quasars \citep{Niida_2020} and 
X-ray selected AGNs \citep[][the data with the shaded region]{Ueda_2014}, for which we utilize the bolometric correction factors from \citet{Richards_2006} and \cite{Duras_2020}, respectively.
For rest-optical selected AGNs through JWST imaging and spectroscopic observations, we present the abundance of broad-line AGNs \citep{Maiolino_2023_JADES,Harikane_2023_agn} and dust-reddened AGNs \citep{Greene_2024, Kokorev_2024, Akins_2024LRD}. We note that these bolometric luminosities taken from literatures have already been corrected for dust attenuation.

Our AGN candidates (red), characterized by dual-line emission and compactness, show an abundance of 
$\Phi_{\rm bol} \sim 10^{-4}-10^{-3}~\mpc^{-3}~{\rm dex}^{-1}$ across luminosities 
$3\times 10^{43}\lesssim L_{\rm bol}/({\rm erg~s}^{-1}) \lesssim 10^{45}$.
This abundance significantly exceeds expectations from UV and X-ray selected AGN populations \citep{Ueda_2014,Niida_2020},
but agrees well with those of unobscured faint broad-line AGNs \citep{Maiolino_2023_JADES,Harikane_2023_agn}, thus suggesting that a significant fraction of our dual-line emitters also host broad Balmer lines.

We model the binned bolometric LF by fitting the data with a DPL function
\begin{equation}
\Phi(L_{\rm bol})=
\dfrac{\Phi^\ast_{\rm bol}}{(L_{\rm bol}/L_{\rm bol}^\ast)^{\bar{\alpha}} + (L_{\rm bol}/L_{\rm bol}^\ast)^{\bar{\beta}}}
\end{equation}
where $\Phi^\ast_{\rm bol}$ (in units of Mpc$^{-3}$ dex$^{-1}$) is the overall normalization, 
$L_{\rm bol}^\ast$ is the characteristic luminosity, $\bar{\alpha}$ and $\bar{\beta}$ are the faint- and bright-end slopes, respectively. 
Similar to the UV LF, we apply both our AGN candidates and unobscured AGNs from ground-base surveys.
The AGN bolometric luminosity function can be fitted by a DPL form with $\Bar{\alpha}=0.62\pm0.12$
and $\bar{\beta}=1.72\pm0.12$ at the faint and bright ends, respectively, and the characteristic break of the LF occurs at 
$L_{\rm bol}^\ast \simeq 3.2\times 10^{44}~{\rm erg~s}^{-1}$ (see Table~\ref{tab:fit}).

%%%%%%%%
%	Fig.11    %
%%%%%%%%
\begin{figure}
\begin{center}
{\includegraphics[width=83mm]{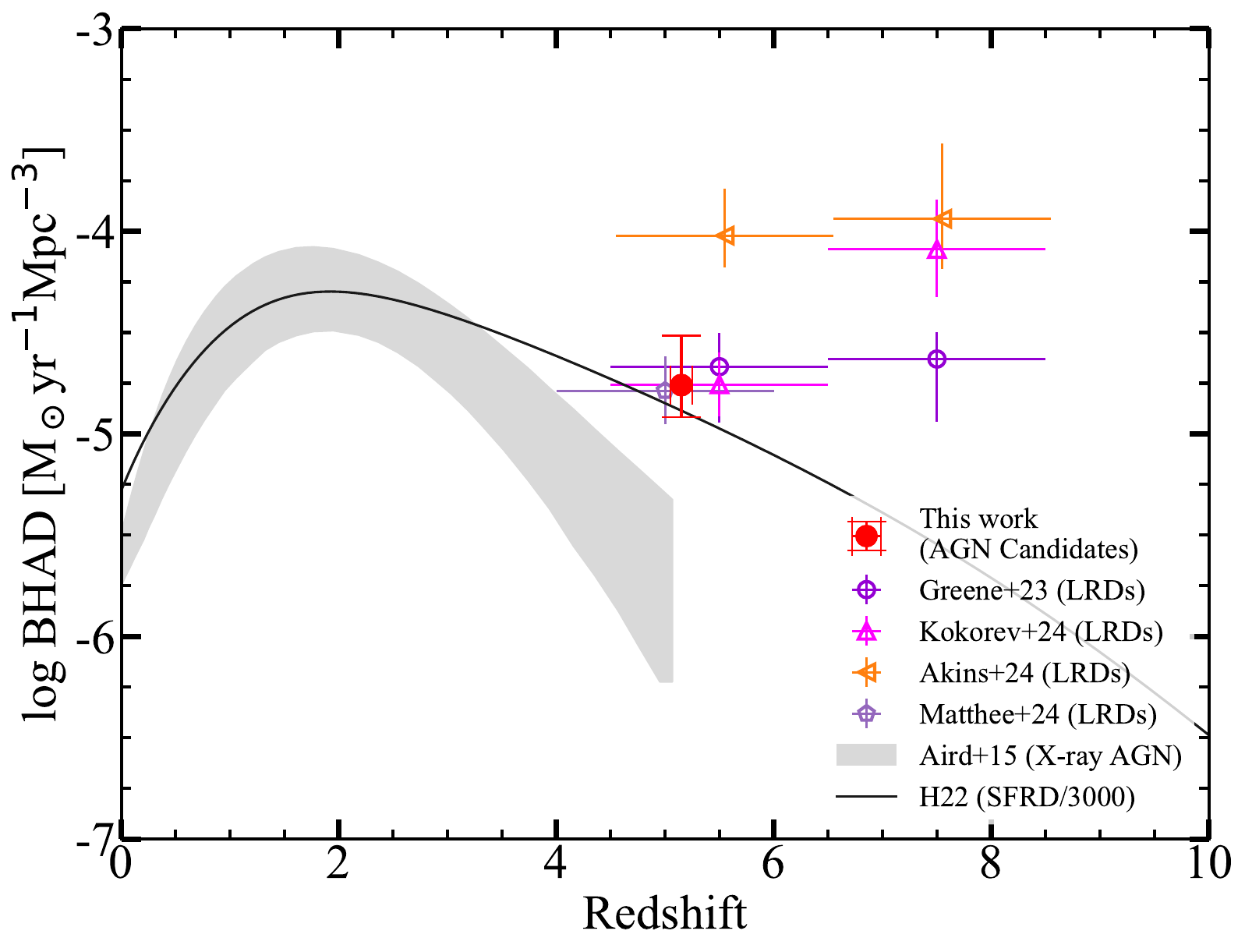}}	
\caption{The BHAD as a function of redshift. Each data point and curve represent BHADs estimated under the 
assumption of a 10\% radiative efficiency ($\epsilon_{\rm rad} = 0.1$) for different types of AGNs, 
including our unobscured AGN candidates (red), LRDs \citep{Matthee_2024,Greene_2024,Kokorev_2024,Akins_2024LRD}, 
and X-ray selected AGNs \citep[gray,][]{Aird_2015}.
For comparison, the cosmic SFRD scaled by a factor of 3,000 is overlaid \citep{Harikane_2022a}.
The BHAD calculated from the JWST-identified AGNs remains significantly dominant at $z>5$.
}
\label{fig:BHAD}
\end{center}
\end{figure}

Furthermore, we calculate the BH accretion rate density (BHAD) by integrating the best-fit bolometric LF as
\begin{align}
\dot{\rho}_{\bullet}&=\frac{1-\epsilon_{\rm rad}}{c^2 \epsilon_{\rm rad}}
\int L_{\rm bol}\Phi(L_{\rm bol}) ~{\rm d}\log L_{\rm bol},
\nonumber\\[4pt]
&\simeq (1.74_{-0.52}^{+1.32}) \times 10^{-5}~\msunyr ~ \mpc^{-3},
\end{align}
where $\epsilon_{\rm rad}=0.1$ is the radiative efficiency of accreting material onto BHs,
and the integration is performed over $43 \leq \log [L_{\mathrm{bol}}/({\rm erg~s}^{-1})] \leq 47$.
In Figure~\ref{fig:BHAD}, we present the BHAD across various redshifts for different types of AGNs,
including our unobscured AGN candidates (red), LRDs \citep{Matthee_2024,Greene_2024,Kokorev_2024,Akins_2024LRD}, 
and X-ray selected AGNs \citep[gray,][]{Aird_2015}.
For each data point and curve, the BHAD is estimated by assuming a canonical 10\% radiative efficiency for accreting BHs 
\citep[see][]{Soltan_1982,Yu_Tremaine_2002}.
From the nearby to high-redshift universe, the BHAD estimated from X-ray sources peaks at $z\simeq 2$ and decreases.
At $0\lesssim z\lesssim 3$, the BHAD follows a similar evolution to the cosmic star-formation rate density (SFRD) 
scaled by a factor of 3,000 \citep{Harikane_2022a}, as implied by BH-galaxy coevolution scenarios \citep{Kormendy_Ho_2013}. 
Toward higher redshifts ($4 < z < 8.5$), the BHAD inferred from the JWST-identified AGNs indicates a persistent or even 
increasing trend, in contrast to the declining trend for X-ray selected AGNs.
The result for our unobscured AGN candidates is fairly consistent with those for dust-reddened broad-line AGNs
\citep{Matthee_2024,Greene_2024}, but is approximately five times lower than that of LRDs observed by 
the COSMOS-Web survey, which covers wide areas and thus can identify rarer but brighter LRD sources \citep{Akins_2024LRD}.
This discrepancy arises from the contribution from brighter LRDs to the BHAD at $L_{\rm bol}\gtrsim 10^{45}~{\rm erg~s}^{-1}$,
where the luminosity function of LRDs extends with a flat slope as $\Phi (L_{\rm bol}) \propto L_{\rm bol}^{-0.71}$
(see the dotted line of Figure~\ref{fig:Lbol}).

\subsection{H$\alpha$ and [O~{\sc iii}] luminosity functions}\label{sec:LFO3Ha}

%%%%%%%%
%	Fig.12.     %
%%%%%%%%
\begin{figure*}
\begin{center}
{\includegraphics[width=85mm]{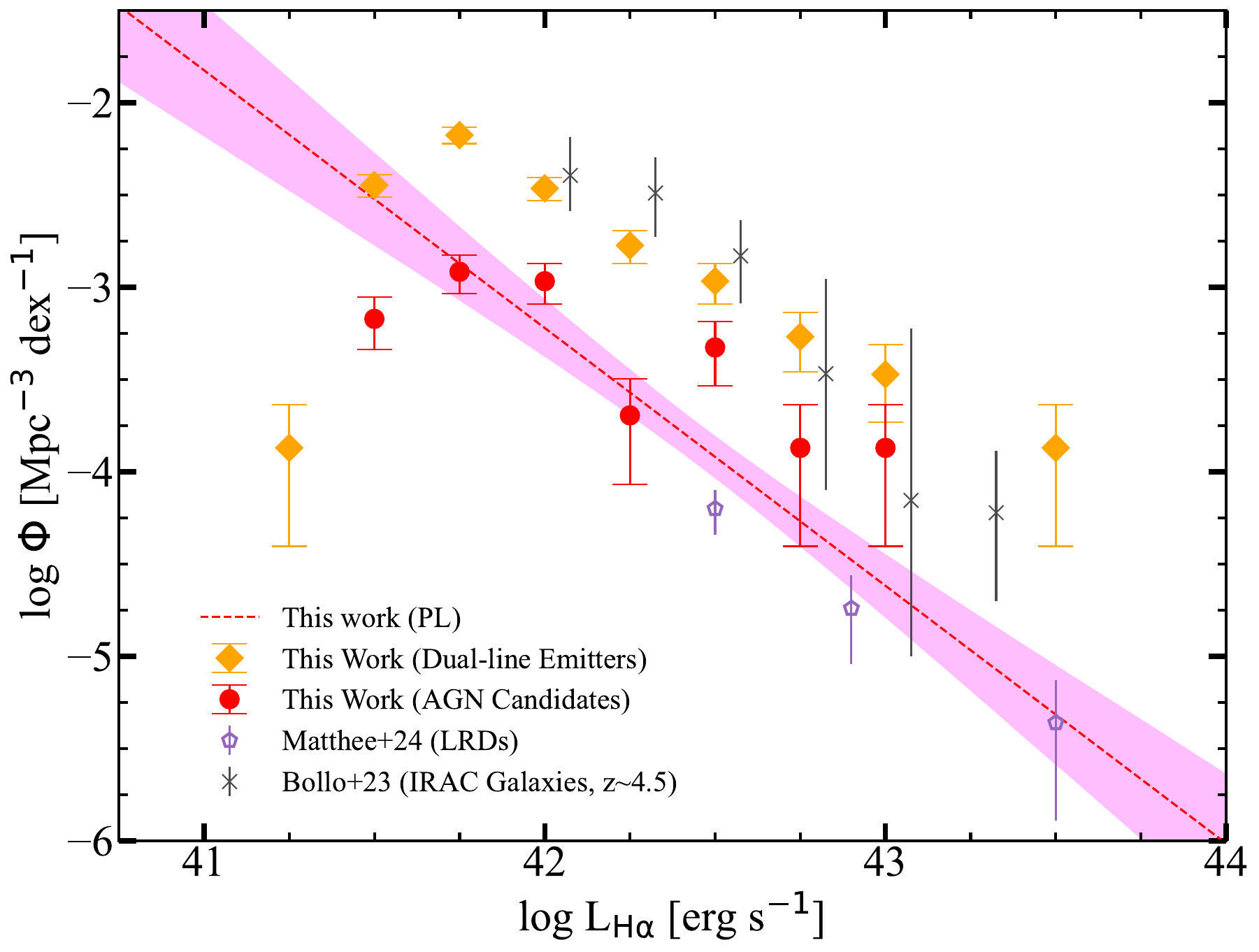}}\hspace{3mm}
{\includegraphics[width=85mm]{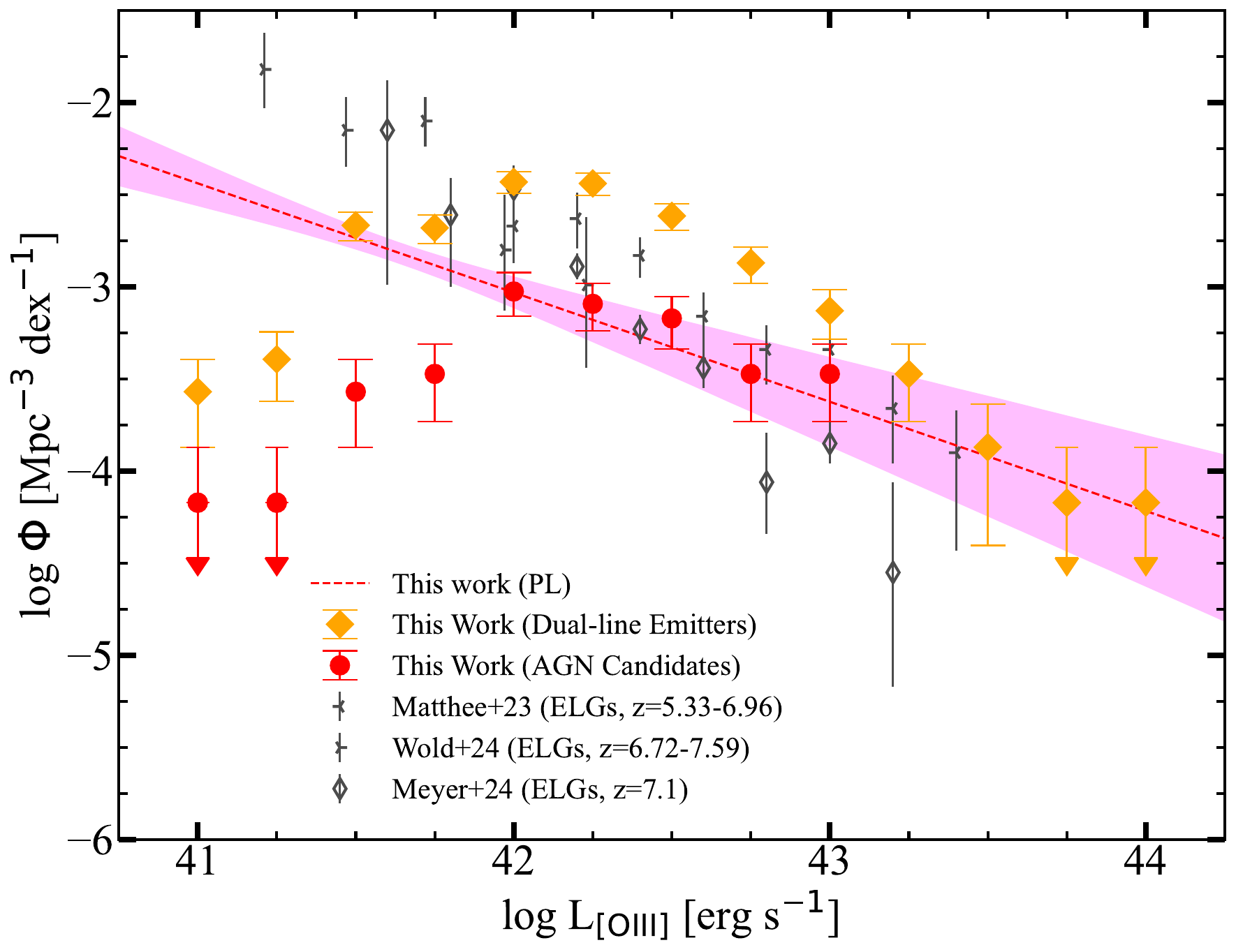}}
\caption{The H$\alpha$ (left) and [\ion{O}{3}] (right) luminosity functions of dual-line emitting galaxies (orange) 
and unobscured AGNs (red).
For comparison, Our estimation of all dual-line emitters are consistent with  the samples from \citet{Bollo_2023}, which are investigated by Spizer/IRAC and presented with black crossing. We also show the H$\alpha$ LF of LRDs from \citet{Matthee_2024} with purple pentagon.
For [\ion{O}{3}] LF, we overplot the result of emission line galaxies (ELGs) from \citet{Matthee_2023_EIGERII}, \citet{ Meyer_2024} and \citet{Wold_24_OIIILF}, which focus on slightly higher redshifts.
}
\label{fig:LHa}
\end{center}
\end{figure*}

Given our targets selected by the presence of strong H$\alpha$ and \HbOIII emission lines, 
we investigate the luminosity functions of these emission lines. 
Utilizing the rest-frame equivalent width for each line calculated earlier in Section~\ref{sec:LineProp}, 
we convert these widths into the line luminosities.
The luminosities for H$\alpha$ and H$\beta$+[\ion{O}{3}] are derived from the observed flux excess in the F410M and F277W bands,
respectively, over the baseline continuum flux.
To extract the [O~{\sc iii}] luminosity, we deduce the H$\beta$ contribution, adopting a ratio of 
$L_{\rm H\beta} = L_{\rm H\alpha}/3.1$ as suggested in \cite{Osterbrock_1989}.

Figure~\ref{fig:LHa} presents the H$\alpha$ (left) and [O~{\sc iii}] (right) luminosity functions for 
dual-line emitting galaxies (orange) and unobscured AGN candidates (red).
These functions cover a wide range of the luminosities 
over $10^{41}\lesssim L_{\rm H\alpha}(L_{\rm [O III]})/{\rm erg~s}^{-1} \lesssim 3\times 10^{43}$,
with a decline in abundance at $\lesssim 10^{42}~{\rm erg~s}^{-1}$ due to sample incompleteness.
For comparison among galaxy populations, we overlay the H$\alpha$ luminosity function of $z\sim 4.5$ Lyman break 
galaxies that exhibit color excess in the Spizer/IRAC $3.6~\mu$m and $4.5~\mu$m bands \citep[][grey cross]{Bollo_2023}, 
as well as the [\ion{O}{3}] luminosity function of star-forming galaxies at $z\simeq 5 - 7$ identified through JWST slitless 
spectroscopy modes \citep{Matthee_2023_EIGERII}.
Remarkably, the luminosity functions of our dual-line emitting galaxies align with those measured in the other studies
using different selection techniques.
This consistency further validates the robustness of our sample and its effectiveness in representing 
the overall galaxy population, as illustrated in Figure~\ref{fig:UVLF}.

Moreover, the H$\alpha$ LF for our unobscured AGN candidates with $L_{\rm H\alpha}\sim (0.3-1.0)\times 10^{43}~{\rm erg~s}^{-1}$
is generally consistent with the observed abundance of AGNs with broad H$\alpha$ emission \citep{Matthee_2024}.
However, it is important to note that the AGN sample in \cite{Matthee_2024} are considered to be significantly affected by
dust extinction \citep[e.g.,][]{Labbe_2023,Greene_2024,Li_LRD_2024}. 
This apparent abundance match between our unobscured AGN candidates and LRDs would imply that
the two different populations are essentially similar but reside in different environments,
influenced by the presence or absence of dense dusty medium.
Fitting both datasets with a single power-law form, the H$\alpha$ luminosity function is characterized by
$\Phi (L_{\rm H\alpha}) \propto L_{\rm H\alpha}^{-1.39}$, which is steeper than the slope of the bolometric LF
for JWST-identified AGNs (dotted curve of Figure~\ref{fig:Lbol}).

\subsection{BH Mass function}

\begin{figure}
\begin{center}
{\includegraphics[width=83mm]{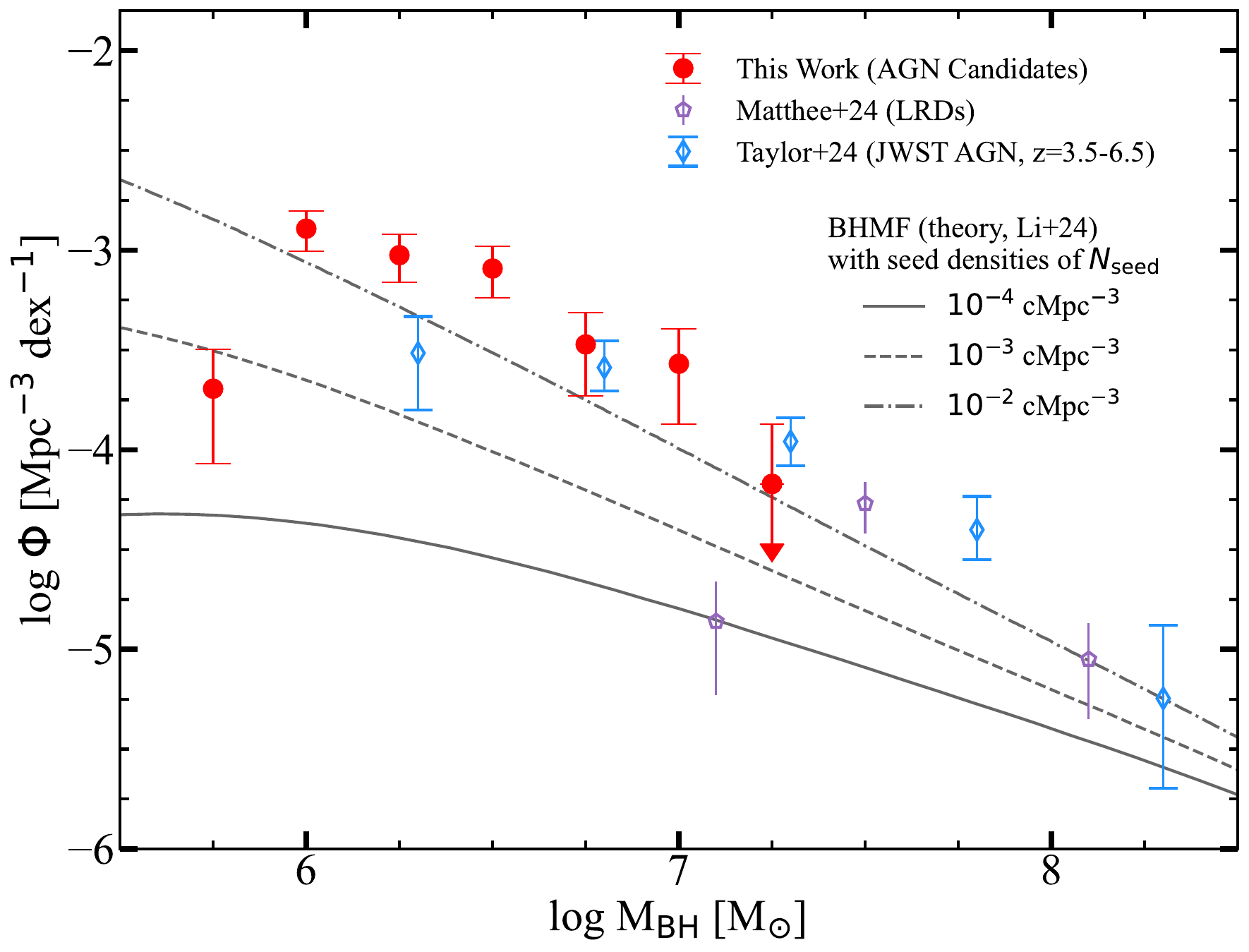}}
\caption{The $z\sim5$ BHMF. Here we show the BHMF of our AGN candidates assuming a constant Eddington ratio $\lambda_{\mathrm{Edd}}=0.26$, which is the median value of \citet{Maiolino_2023_JADES}. 
For comparison, we show the BHMF of LRDs from \citet{Matthee_2024} with purple pentagons and that of JWST broad-line AGNs from \citet{Taylor_24_BLAGN} with blue diamonds that both cover higher mass ranges.
We also overplot the simulation result from \citet{Li_LF_2024}. Our estimation is consistent with the simulated BHMF assuming a seed density of $N_{\mathrm{seed}}=10^{-2}$ Mpc$^{-3}$.}
\label{fig:BHMF}
\end{center}
\end{figure}

Here, we discuss the shape of the BH mass function (BHMF) at $z\sim5$ derived from the bolometric luminosity 
function of our AGN candidates (see Figure~\ref{fig:Lbol}).  
Since no direct BH mass measurements are conducted in this work, we instead estimate the BH mass by assuming a 
constant Eddington ratio. 
For this analysis, we adopt an Eddington ratio of $\langle \lambda_{\rm Edd}\rangle \simeq 0.26$, the median value of 
the unobscured AGN samples observed with JWST \citep{Maiolino_2023_JADES}.
Therefore, the BH mass is approximated as 
\begin{equation}
    M_{\rm BH}\simeq 3.0\times 10^6~\msun \left(\frac{L_{\rm bol}}{10^{44}~{\rm erg~s}^{-1}}\right)
    \left(\frac{\langle \lambda_{\rm Edd}\rangle }{0.26}\right)^{-1}.
\end{equation}
Under this assumption, we construct the BHMF at the redshift of $z=5.03-5.26$ as shown in Figure~\ref{fig:BHMF}.
The masses are distributed in the range of $M_{\rm BH}\simeq 10^6-10^7~\msun$, with an abundance of 
$\Phi_\bullet \sim 10^{-3.5} - 10^{-3}~{\rm Mpc}^{-3}~{\rm dex}^{-1}$.

Compared with the BHMF over $10^{7.5}<M_{\rm BH}/\msun < 10^8$ constructed by \citet{Matthee_2024},
our AGN samples are distributed in a mass range $\sim 1$ dex lower and have a number density $\sim 1$ dex higher. 
Our BHMF is also consistent with that from \citet{Taylor_24_BLAGN}, which is constructed by JWST spectroscopically confirmed broad-line AGN with a larger survey volume. 
We find agreement with the predicted BHMF at the low-mass regime by \citet{Li_LF_2024}, 
which is the result of a semianalytical model for BH formation and growth that considers multiple accretion bursts 
with variable Eddington ratios and is primarily calibrated with the quasar abundance at the bright end of $\Muv<-24$
(see also comparison to the UV luminosity function in \citealt{Kocevski_2024}).

A quantitative comparison suggests that our data aligns with a model featuring a seed BH density of
$N_{\rm seed}\simeq 10^{-2}~\mpc^{-3}$, which represents an efficient seed formation scenario compared to
the typical case with $N_{\rm seed}\simeq 10^{-3}~\mpc^{-3}$ in the progenitor halos of high-$z$ quasar host galaxies.
Further spectroscopic observations with JWST/NIRSpec will provide more precise measurements of the BHMF at the low-mass regime,
allowing for robust constraints on the number density of the initial seed BHs formed at $z>10$.

%%%%%%%%
%	Fig.13     %
%%%%%%%%

% \subsection{Spectroscopy Confirmation in the Future}

% We plan to conduct follow-up observations of our AGN candidates to spectroscopically confirm 
% the presence of broad-emission lines, in an approved JWST Cycle 3 program 
% (GO\#5718; PI: D. Kocevski, co-PI: J. Guo). 
% This small-size program has been allocated 20.6 hours of assigned spacecraft time. 
% According to the current MSA design, we expect to observe $\gtrsim 20$ of our unobscured AGN candidates within 6 pointings using the F290LP/G395M configuration. 
% With a spectral resolution of $R \sim 1000$ and $\simeq 2$ hours of exposure time for each object,
% we will able to measure the masses of the BHs powering those unobscured AGNs down to $M_{\rm BH} \sim 10^6~\msun$ with significant confidence.

\section{Summary} \label{sec:conclusion}

In this work, we search for $z=5$ galaxies in the JWST CEERS field that emit strong \HbOIII and H$\alpha$ lines, traced by photometric excess in NIRCam's F277W and F410M filters.
Our photometric selection requiring these two emission lines is efficient in selecting galaxies within a narrow range of $5.03\leq z\leq 5.26$, as has been demonstrated in the early discovery of CEERS 2782, a $z=5.24$ AGN, by  \citet{Onoue_2023} and \citet{Kocevski_2023}.

% The JWST has enabled us to uncover faint galaxies and AGNs in the early universe. 
% Taking advantage of the unique filter combination used in the CEERS program, we perform an extensive photometric search of galaxies emitting strong \HbOIII and H$\alpha$ lines.
% By requiring photometric excess in NIRCam's F277W and F410M images, the redshift range of the galaxies is limited to $5.03\leq z\leq 5.26$.
% This dual-line emitter selection is motivated by early high-redshift AGN studies of \cite{Onoue_2023} and \cite{Kocevski_2023}, where they discovered a $z=5.24$ AGN, 
% CEERS 2782, which shows emission line excesses of \HbOIII in F277W and H$\alpha$ in F410M.

A total of 261 \HbOIII and H$\alpha$ dual-line emitters are selected over absolute UV magnitudes of $-22\lesssim M_{\mathrm{UV}}\lesssim -17$.
The mean rest-frame equivalent width of these galaxies is 1010 {\AA} for \HbOIII and 1040 {\AA} for H$\alpha$, consistent with that of CEERS 2782.
Their UV luminosity function is well described by the Schechter function model of $z=5$ LBGs, when scaled down by a factor of 0.4. 
Thus, the dual-line emitters in this study represent a substantial fraction of the UV-selected galaxy population, despite the different selection criteria.

% Our extensive search identifies a total of 261 \HbOIII and H$\alpha$ dual-line emitters over absolute UV magnitudes of $-22\lesssim M_{\mathrm{UV}}\lesssim -17$.
% The mean values of rest-frame equivalent widths for \HbOIII and H$\alpha$ are 1010 {\AA} and 1040 {\AA}, respectively, consistent with those observed in CEERS 2782.
% The UV luminosity function of the dual-line emitting galaxies is well described with a Schechter function, and this population accounts for $\sim 40\%$ of the Lyman break 
% galaxies at the same redshift range. 
% Therefore, dual-line emitters in this study represent a substantial fraction of the UV-selected galaxy population, despite the different selection criteria.

Based on the size measurements with {\sc Galfit}, we find 58 dual-line emitters (22\% of the whole sample) that exhibit compact morphology at the rest-UV or optical wavelength. 
With an assumption that these compact sources are dominated by AGN, their AGN bolometric luminosity falls in the range of 
$2\times 10^{43} \lesssim L_{\rm bol}/({\rm erg~s}^{-1})\lesssim 3\times 10^{44}$.
Their cosmic abundance ($\Phi \simeq 10^{-3}~{\rm Mpc}^{-3}~{\rm dex}^{-1}$) is two orders of magnitude higher than that  extrapolated from the UV-selected luminous quasars 
\citep[e.g.,][]{Niida_2020}, which is consistent with previous JWST studies of broad-line AGNs \citep[e.g.,][]{Harikane_2023_agn,Maiolino_2023_JADES}.
This high AGN abundance implies an AGN duty cycle of $\lesssim 10\%$ in galaxies at the redshift.

Based on the photometric excess, we measure the H$\alpha$ and [\ion{O}{3}] luminosity functions of the dual-line emitters down to $\lesssim 10^{42}~{\rm erg~s}^{-1}$. 
Our results based on the photometric sample align with those measured in other JWST studies using different selection techniques \citep{Bollo_2023,Matthee_2023_EIGERII,Matthee_2024}.
Moreover, the H$\alpha$ luminosity function of the unobscured AGN candidates is broadly consistent with that of LRDs that cover the luminous end.
% closely matches that of LRDs with broad H$\alpha$ emission,
% despite differences both in selection criteria and population characteristics.

% Moreover, our dual-line emitter sample reaches the faint end of the H$\alpha$ and [\ion{O}{3}] luminosity functions down to $\lesssim 10^{42}~{\rm erg~s}^{-1}$. 
% The luminosity functions of our samples align with those measured in the other studies using different selection techniques \citep{Bollo_2023,Matthee_2023_EIGERII,Matthee_2024}.
% In particular, the H$\alpha$ luminosity function of our unobscured AGN candidates closely matches that of LRDs with broad H$\alpha$ emission,
% despite differences both in selection criteria and population characteristics.

Spectroscopic follow-up observations of these dual-line emitters are planned in an approved JWST Cycle~3 program (GO\#5718; PI: D. Kocevski, co-PI: J. Guo), in which we will use the NIRSpec Micro-Shutter Assembly to confirm their nature.
For the compact unobscured AGN candidates, we aim to characterize their BH activity and construct their mass distribution down to $M_{\rm BH} \sim 10^6~\msun$, thereby improving our understanding of the BH seeding mechanisms and their growth history at cosmic dawn.

\begin{acknowledgments}
We greatly thank Vasily Kokorev, Zhengrong Li, Hanpu Liu, \'Oscar A. Ch\'avez Ortiz, and Anthony J. Taylor for constructive discussions. 
K.~I. and M.~O. acknowledge support from the National Natural Science Foundation of China (12073003, 12003003, 11721303, 11991052, 11950410493), 
and the China Manned Space Project (CMS-CSST-2021-A04 and CMS-CSST-2021-A06). 
J.~G. acknowledges the support from the Undergraduate Research Project in Peking University.
M.~O. acknowledges support from the Japan Society for the Promotion of Science (JSPS) KAKENHI grant No. JP24K22894.
The data products presented herein were retrieved from the Dawn JWST Archive (DJA). DJA is an initiative of the Cosmic Dawn Center, which is funded by the Danish National Research Foundation under grant DNRF140.
We wish to thank the entire JWST team and the CEERS collaboration for the operation of the telescope and for developing their observing program with a zero-exclusive-access period. This work is based on observations made with the NASA/ESA/CSA James Webb Space Telescope and these observations are associated with program 1345.

\end{acknowledgments}

\vspace{5mm}
\facilities{JWST,HST}

\software{astropy \citep{2013A&A...558A..33A}, Galfit \citep{Peng_2010_Galfit}}

\bibliography{ref}
\bibliographystyle{aasjournal}

\appendix

\section{The Source Catalog of $z=5$ Dual-Line Emitters}
Here, we present the catalog of the $z=5$ dual-line emitters from this study in Table~\ref{tab:catalog}.

\begin{table}[!ht]
\caption{Dual-Line Emitters}
    \centering
    \begin{tabular}{rccccccclll}
    \hline\hline
        ID   & R.A.       & Dec.      & M.& C. & F356W          & $\alpha_\lambda$ & M$_{\mathrm{UV}}$ & EW$_{\mathrm{H\beta+[O III]}}$ [\AA]& EW$_{\mathrm{H\alpha}}$ [\AA]\\ \hline
        515  & 215.106678 & 52.925811 & 0 & ~  & $28.44\pm0.14$ & $-1.68\pm0.25$ & $-17.50\pm0.21$ & $337\pm167$ & $562 \pm216$ \\ 
        1485 & 215.104809 & 52.927719 & 1 & ~  & $26.07\pm0.02$ & $-1.94\pm0.14$ & $-20.25\pm0.11$ & $1056\pm30 $ & $717 \pm39 $ \\ 
        1836 & 215.104637 & 52.929068 & 0 & ~  & $26.75\pm0.04$ & $-1.60\pm0.16$ & $-19.04\pm0.14$ & $988 \pm54 $ & $543 \pm66 $ \\ 
        1953 & 215.156838 & 52.966785 & 1 & ~  & $27.99\pm0.12$ & $-1.87\pm0.33$ & $-18.18\pm0.30$ & $689 \pm152$ & $750 \pm205$ \\ 
        2353 & 214.980334 & 52.841505 & 0 & ~  & $27.61\pm0.07$ & $-1.95\pm0.11$ & $-18.74\pm0.09$ & $918 \pm89 $ & $726 \pm118$ \\ \hline
        
    \end{tabular}\label{tab:catalog}
    \tablecomments{
    ~We only show the first five dual-line emitters here. The complete catalog is available online\footnote{\url{https://github.com/JingsongGuo-astro/Dual\_Emitter}}. \\
    Column (1): object ID in the DJA catalog.   
    Column (2) \& (3): Right Ascension and Declination in degree. 
    Column (4): The compactness flag as described in Sec.~\ref{sec:CompMeas}.
    %Morphology label classified in Section \ref{sec:CompMeas}. 
    Label 0 indicates an extended morphology, and label 1 indicates a compact morphology (thus an AGN candidate in this study). % and is classified as AGN candidates.
    Column (5): Comments. 
    Column (6): The 0\arcsec.5-diameter aperture photometry. Only the F356W magnitudes are presented here. The full photometry is available in the online table.    
    %The magnitude in JWST bands with uncertainty. This column repeats for each band in online table. 
    Column (7): The power-law index of the continuum emission $\alpha_\lambda$. 
    Column (8): The absolute UV magnitude, where we assume $z=5.2$ for each object (see Sec.~\ref{sec:ContProp}).    
    %and the absolute UV magnitude with uncertainty estimated in Section \ref{sec:ContProp}. 
    Column (9) \& (10): The rest-frame equivalent width of \HbOIII and H$\alpha$ (see Sec.~\ref{sec:LineProp}). % with uncertainty calculated in Section \ref{sec:LineProp}.
}

\end{table}

\section{NIRCam mediumband pairs to detect dual-line emitters}\label{sec:DisMidBand}

\begin{figure*}
\begin{center}
{\includegraphics[width=85mm]{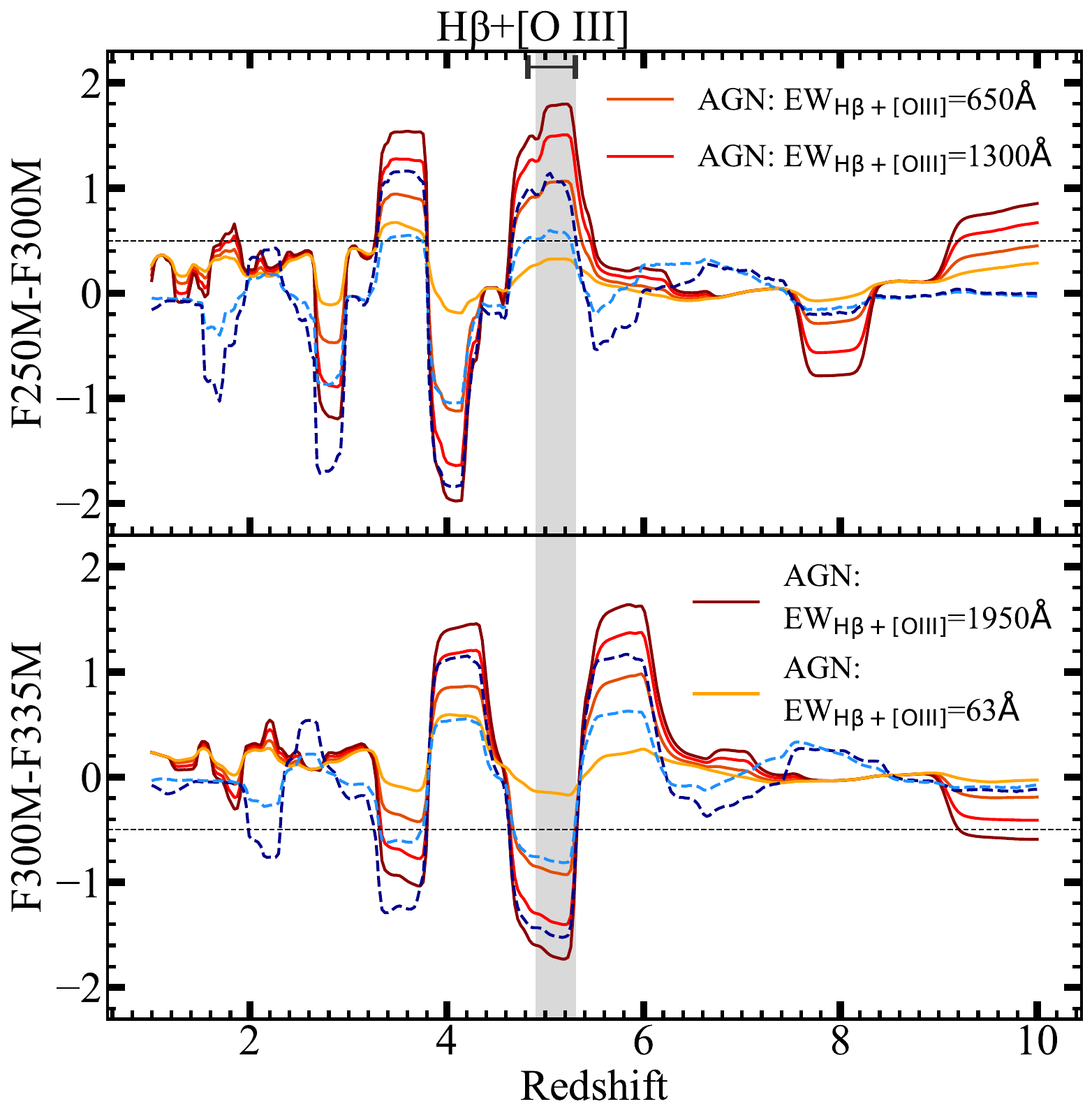}}	
{\includegraphics[width=85mm]{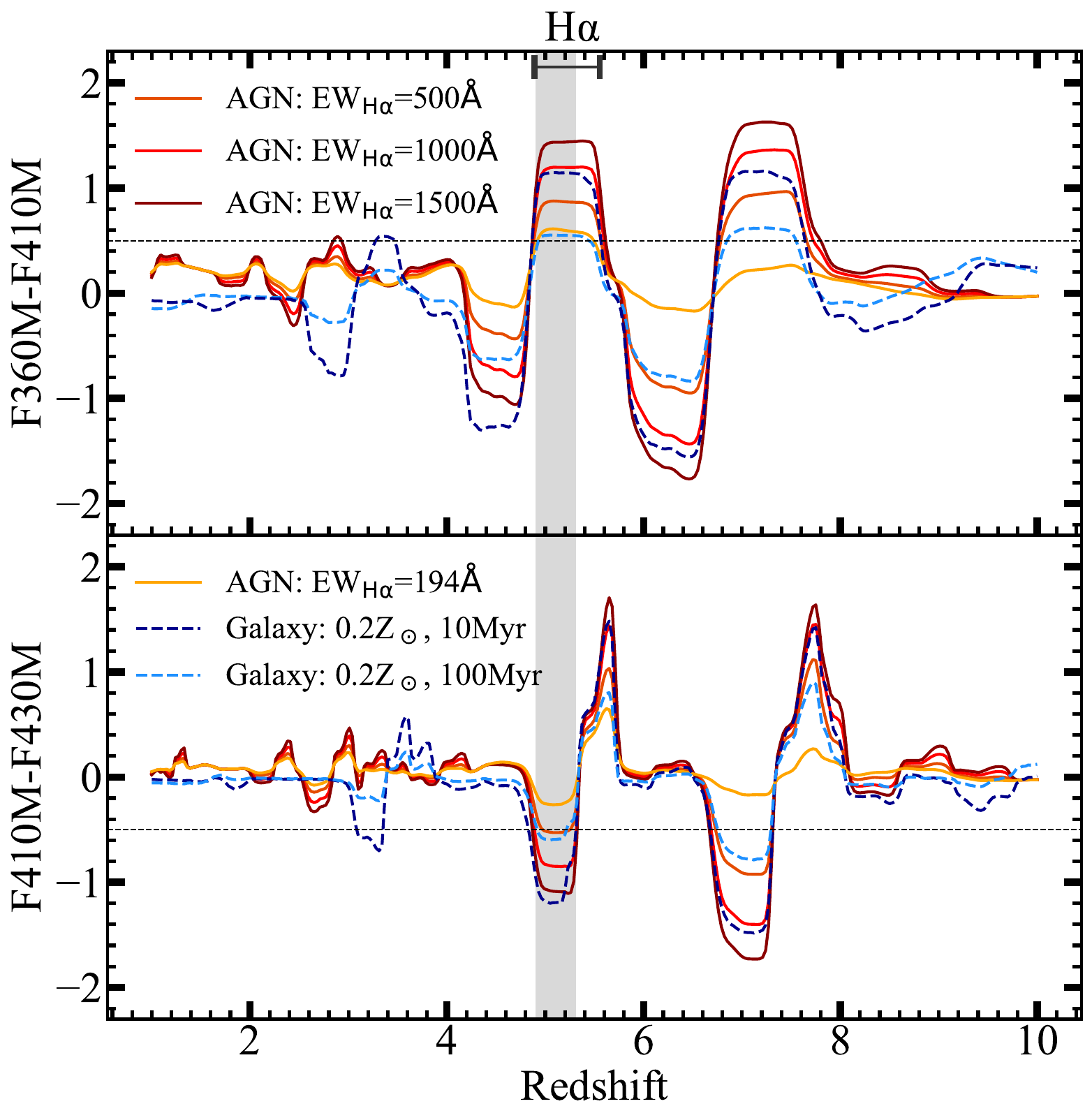}}	
\caption{JWST/NIRCam medium-band colors of AGNs and galaxies as the function of redshift. 
Each curve corresponds to the same AGN and galaxy SED models as shown in Figure~\ref{fig:fig_1_Ha}. 
(\textit{Left:}) F360M $-$ F410M (top) and F410M $-$ F430M (bottom) for capturing a color excess in F410M due to H$\alpha$ emission. (\textit{Right:}) F250M $-$ F300M (top) and F300M $-$ F335M (bottom) for capturing a color excess in F300M due to \HbOIII emission. 
The shaded region indicates the effective redshift range for dual-line emitter search using these color combinations, 
$z=4.90$--$5.31$.}
\label{fig:MidBandEvo}            
\end{center}
\end{figure*}

Here, we explore the potential to find more dual-line emitters at different 
redshifts using additional JWST mediumband filters (F250M, F300M, F335M, F360M, F410M, F430M, F460M, and F480M).
These mediumbands cover almost the entire wavelength range of NIRCam and their bandwidths are narrower than those of broad-bands filters, allowing emission lines to create stronger excesses.
For dual-line emitter search, we focus on two major emission lines, H$\alpha$ and [\ion{O}{3}]$\lambda 5007$, with a wavelength separation of $s=\lambda_{\rm H\alpha}/\lambda_{\rm [OIII]5007}=1.31$. 
This separation allows them to fall in two different medium bands at some specific redshift ranges.
To study this, we simulate the color evolution in different medium bands using the same AGN SED models presented in Section~\ref{sec:Model}, and discuss the useful combinations of medium-band colors to select dual-line emitters. 
While \citet{Withers23} already presents some medium-band pairs for dual-line emitter search at low redshifts (see their Table~1), our focus is on redshifts ranges of $z\gtrsim 5$.

Firstly, we discuss the improvement of our dual-line emitter selection based on the color excess in the F410M filter
due to the H$\alpha$ emission line.
As shown in Section~\ref{sec:excess_Ha}, H$\alpha$ emission falls within the F410M band at redshifts of $z=4.9$--$5.5$. 
However, the effective redshift range of our original selection is reduced to $z=5.03$--$5.26$ (see Sections~\ref{sec:excess_Ha} and \ref{sec:excess_O3}). 
The lower limit is raised because at $z=4.90$--$5.03$ H$\alpha$ enters both in F356W and F410M, resulting in a moderate color excess of F356W $-$ F410M. 
The upper limit is limited by [O{\sc iii}], which enters F356W from $z=5.27$, making the F356W $-$ F410M color less redder than required.
For these reasons, replacing the existing broadband photometry with mediumband images enables us to select galaxies over a wider redshift range.
%at the corresponding wavelengths.

Figure~\ref{fig:MidBandEvo} shows the redshift evolution of the F360M $-$ F410M and F410M $-$ F430M colors (left)
and the F250M $-$ F300M and F300M $-$ F335M colors (right) from $z=1$ to $10$.
Since the wavelength coverage of F360M and F410M is clearly separated, the F360M $-$ F410M color excess is maintained
as long as H$\alpha$ enters the F410M filter ($z=4.9$--$5.5$).
In contrast, the F410M $-$ F430M color becomes sufficiently blue ($\leq -0.5$ mag) only at $z=4.9$--$5.31$, above which H$\alpha$ enters the wavelength range where both the F410M and F430M filters overlap.
It is worth noting that the upper limit of the redshift range using these medium bands is similar to 
that of our original selection, but the underlying reasons are different as describe above.
In this redshift range, the \HbOIII emission lines remain within the F300M band, resulting in sufficiently red and blue colors in F250M $-$ F300M and F300M $-$ F335M, respectively. 
As a result of extending the effective redshift range, the detection number of dual-line emitters is expected to increase by approximately a factor of $\simeq 1.8$.

\begin{figure*}  
\begin{center}
{\includegraphics[width=85mm]{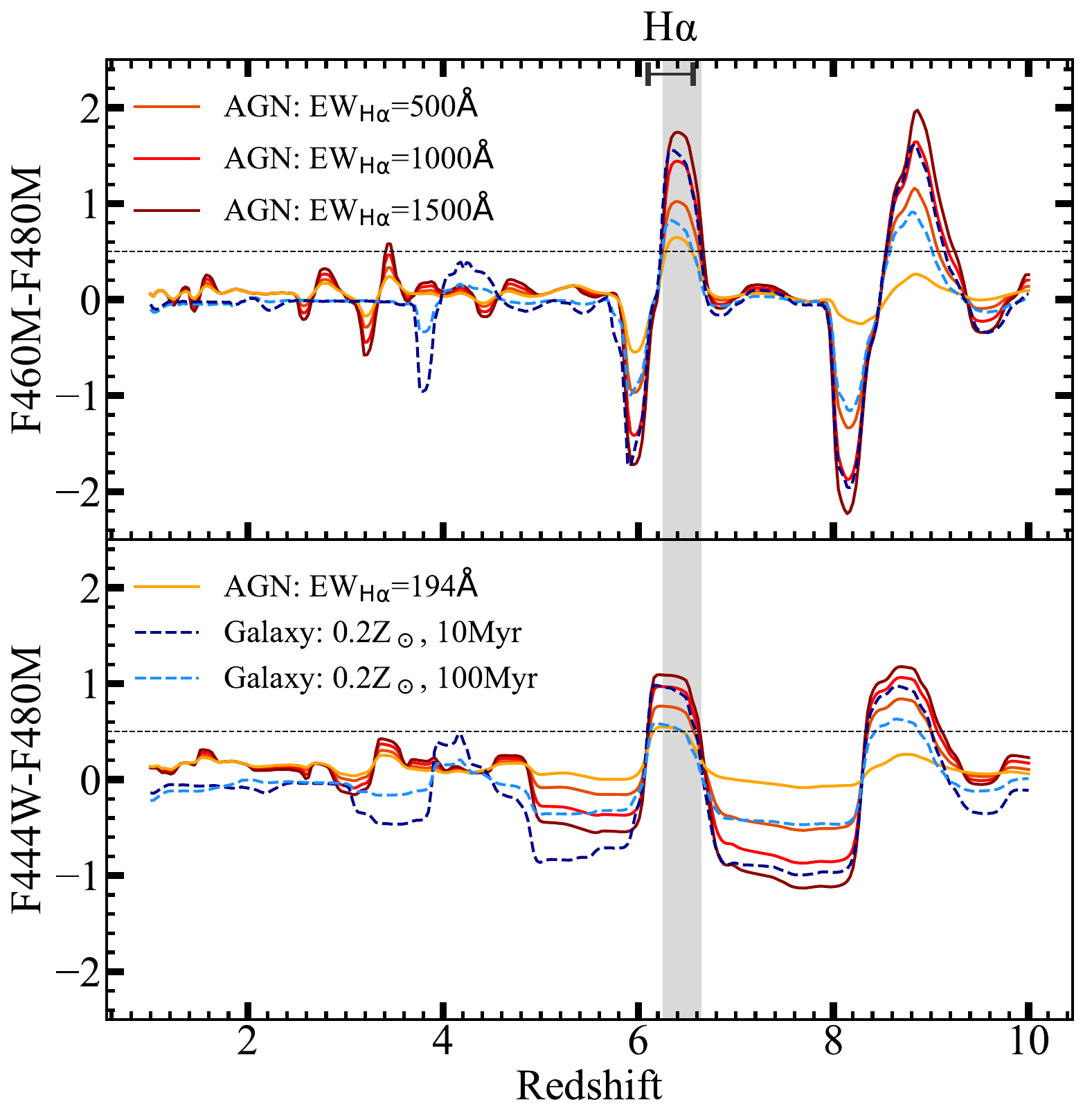}}	
{\includegraphics[width=85mm]{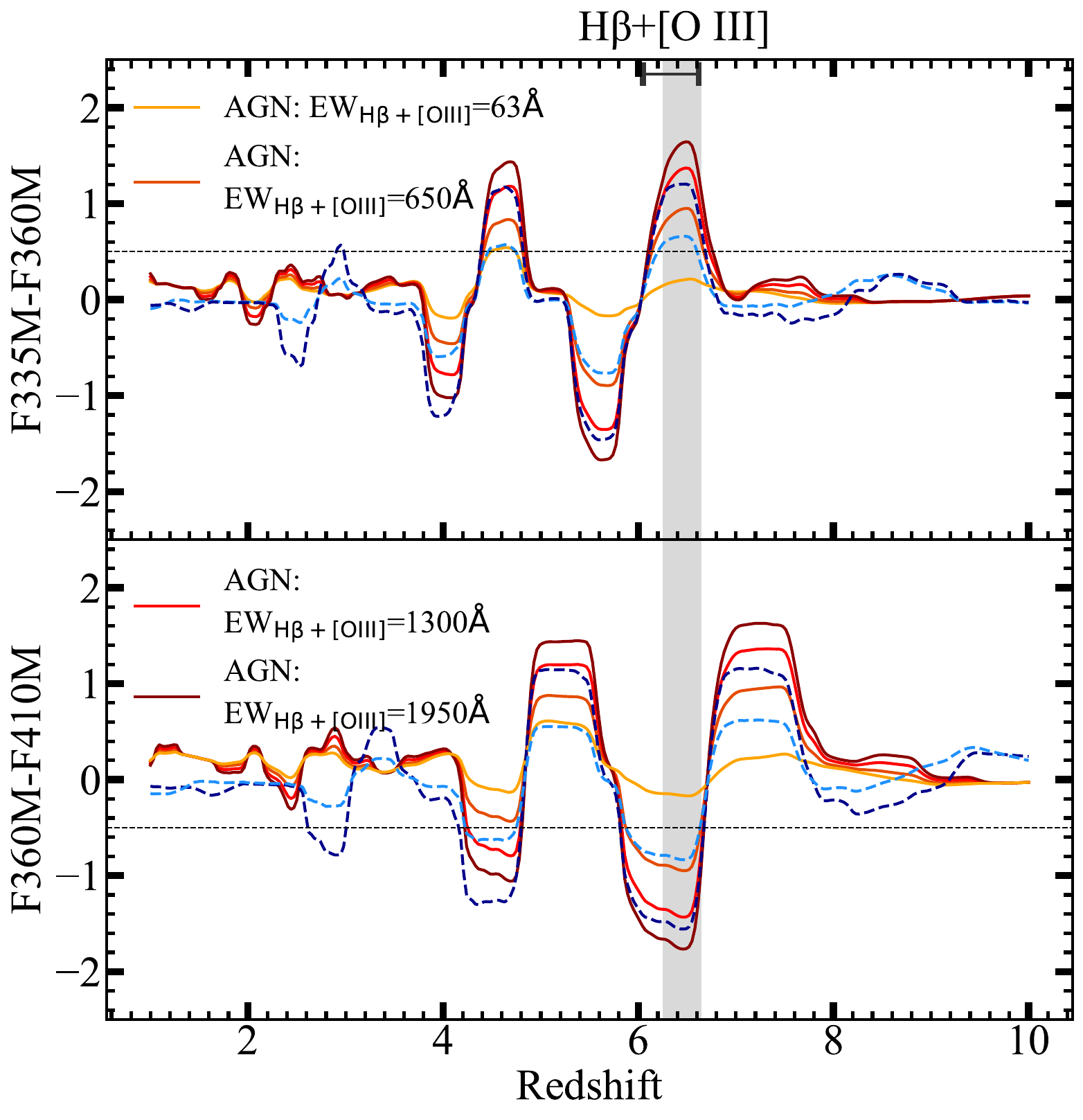}}	
\caption{Same to Figure~\ref{fig:MidBandEvo}, but for different JWST/NIRCam medium-band combinations
to select dual-line emitting galaxies and AGNs at redshifts higher than five.
({\it Left}:) F460M $-$ F480M (top) and F444M $-$ F480M (bottom) for detecting a color excess due to H$\alpha$ emission. (\textit{Right:}) F335M $-$ F360M (top) and F360M $-$ F410M (bottom) for detecting a color excess due to \HbOIII emission. 
The effective redshift range is $z=6.25$--$6.65$, the highest redshift we can explore using NIRCam medium-band filters.
}    
\label{fig:MidBandEvo2}            
\end{center}
\end{figure*}

\begin{figure*}  
\begin{center}
{\includegraphics[width=85mm]{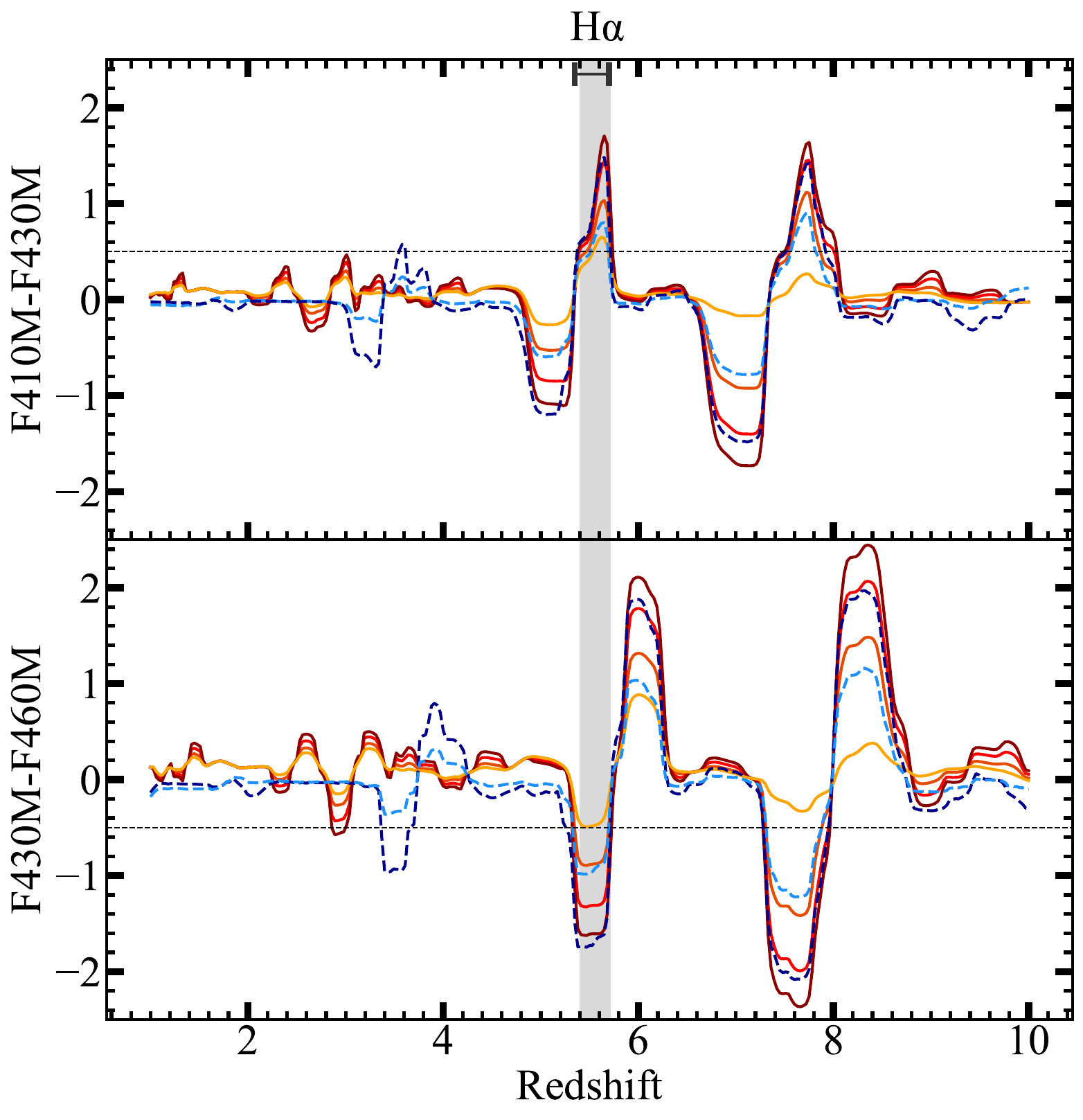}}	
{\includegraphics[width=85mm]{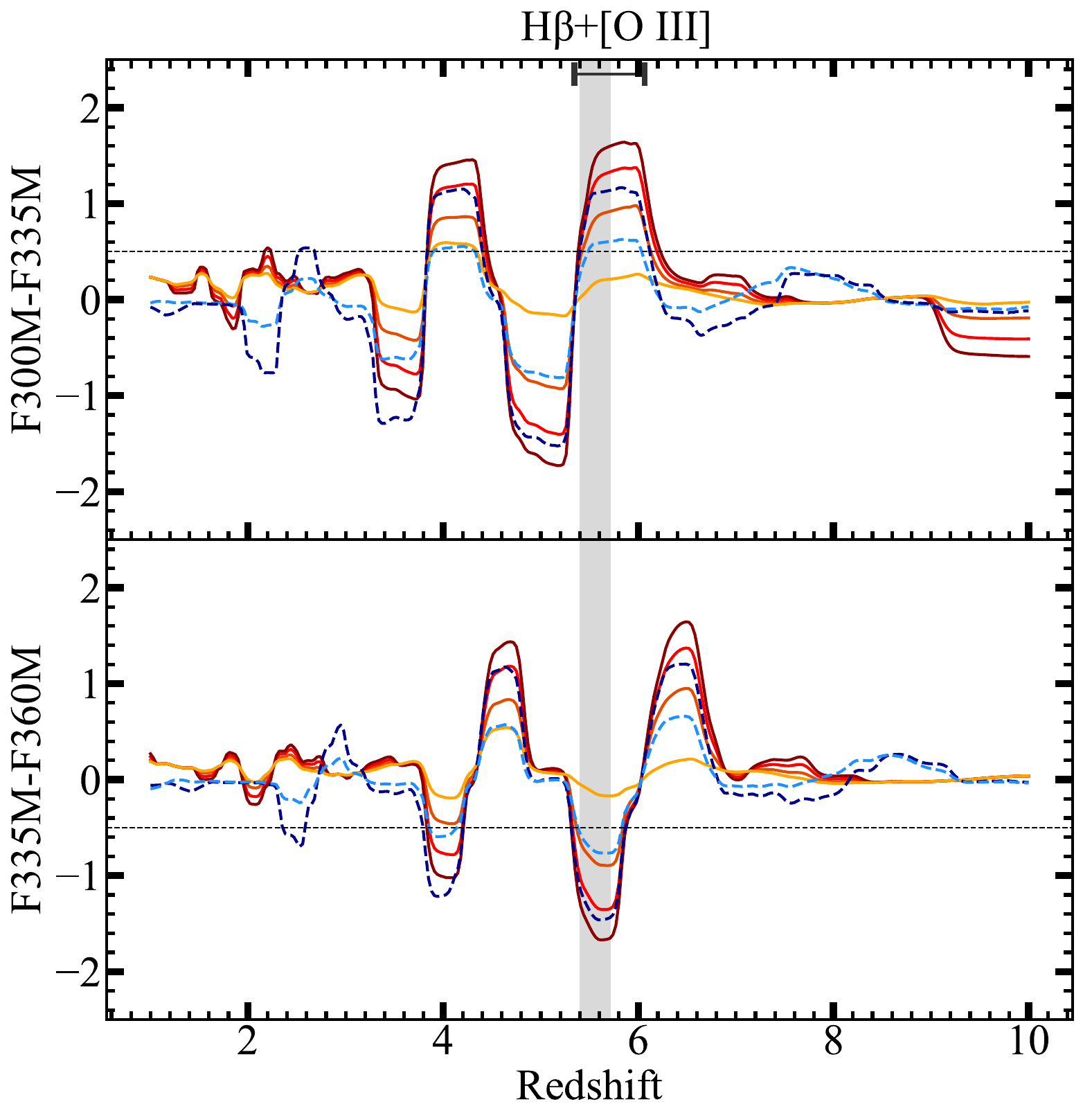}}	
\caption{Same to Figure~\ref{fig:MidBandEvo}, but for different JWST/NIRCam medium-band combinations
to select dual-line emitting galaxies and AGNs at redshifts higher than five.
({\it Left}:) F410M $-$ F430M (top) and F430M $-$ F460M (bottom) for detecting a color excess due to H$\alpha$ emission. (\textit{Right:}) F300M $-$ F335M (top) and F335M $-$ F360M (bottom) for detecting a color excess due to \HbOIII emission. 
The effective redshift range is $z=5.40$--$5.72$.
}    
\label{fig:MidBandEvo3}            
\end{center}
\end{figure*}

Next, we discuss selection criteria for dual-line emitter search at higher redshifts ($z\geq 5.3$) 
using medium-band filters that cover longer wavelengths (e.g., F430M, F460M and F480M).
In Figure~\ref{fig:MidBandEvo2}, we show the color evolution with a filter combination to
search for dual-line emitters at the highest redshift accessible using NIRCam medium bands. 
At the redshift range of $z=6.25$ -- $6.65$, H$\alpha$ is in F480M and \HbOIII are in F360M. 
To detect the H$\alpha$ emission, one can use either the F460M $-$ F480M or F444W $-$ F480M color.
For detecting  the \HbOIII emission lines, the F335M $-$ F360M and F360M $-$ F410M colors can be adopted.

Additionally, the Figure~\ref{fig:MidBandEvo3} present the other combination of medium-band filters,
where H$\alpha$ and \HbOIII enter the F430M and F335M band, respectively.
In this case, the effective redshift range is $z=5.40$--$5.72$ (note that the color excess in the F410M $-$ F430M
is modest due to the emergence of H$\alpha$ both in F410M and F430M).

\end{document}